\documentclass[letterpaper, journal, 10pt]{IEEEtran}

\usepackage{etex}

\usepackage{ifpdf}
\ifpdf
\usepackage[draft,pdfborder={0 0 0}]{hyperref}
\fi

\usepackage{cite}

\usepackage{algorithm}
\usepackage{algorithmicx}
\usepackage{algpseudocode}

\usepackage{listings}

\usepackage[cmex10]{amsmath}
\usepackage{amssymb}
\usepackage{nicefrac}

\usepackage{bm}
\usepackage[varg]{txfonts}
\let\mathbb=\varmathbb
\DeclareSymbolFont{letters}{OML}{ztmcm}{m}{it}

\usepackage[font=footnotesize]{subfig}

\usepackage{tabularx}
\usepackage{multirow}
\usepackage{dcolumn}
\usepackage{booktabs}

\usepackage{tikz}
\usepackage{pgfplots}
\usetikzlibrary{plotmarks,shapes,positioning,arrows,decorations.markings,fit,calc,patterns,backgrounds}
\makeatletter
\tikzset{circle split part fill/.style args={#1,#2}{%
    alias=tmp@name,%
    postaction={%
      insert path={
        \pgfextra{%
          \pgfpointdiff{\pgfpointanchor{\pgf@node@name}{center}}%
          {\pgfpointanchor{\pgf@node@name}{east}}%
          \pgfmathsetmacro\insiderad{\pgf@x}
          \fill[#1] (\pgf@node@name.base) ([xshift=-\pgflinewidth]\pgf@node@name.east) arc
          (0:180:\insiderad-\pgflinewidth)--cycle;
          \fill[#2] (\pgf@node@name.base) ([xshift=\pgflinewidth]\pgf@node@name.west)  arc
          (180:360:\insiderad-\pgflinewidth)--cycle;
        }}}}}
\makeatother  

\usepackage{siunitx} 
\usepackage[T1]{fontenc} 

\newcommand{\tparam}[1]{\textless#1\textgreater}
\newcommand{\sgn}[1]{\text{sgn}(#1)}

\newcommand{\mvec}[1]{\bm{#1}}

\newcolumntype{C}[1]{>{\centering}m{#1}}

\begin{document}

\title{Low-Latency Software Polar Decoders}

\author{Pascal Giard, Gabi Sarkis, Camille Leroux, Claude Thibeault, and Warren J. Gross%
\thanks{P. Giard, G. Sarkis and W. J. Gross are with the Department of Electrical and Computer Engineering, McGill University, Montr\'eal, Qu\'ebec, Canada (e-mail: \{pascal.giard,gabi.sarkis\}@mail.mcgill.ca, warren.gross@mcgill.ca).}%
\thanks{C. Leroux is with the IMS Lab, Bordeaux-INP, Bordeaux, France (e-mail: camille.leroux@ims-bordeaux.fr).}%
\thanks{C. Thibeault is with the Department of Electrical Engineering, \'Ecole de technologie sup\'erieure, Montr\'eal, Qu\'ebec, Canada (e-mail: claude.thibeault@etsmtl.ca).}}

\maketitle

\begin{abstract}
Polar codes are a new class of capacity-achieving error-correcting codes with low encoding and decoding complexity. Their low-complexity decoding algorithms rendering them attractive for use in software-defined radio applications where computational resources are limited. In this work, we present low-latency software polar decoders that exploit modern processor capabilities. We show how adapting the algorithm at various levels can lead to significant improvements in latency and throughput, yielding polar decoders that are suitable for high-performance software-defined radio applications on modern desktop processors and embed\-ded-plat\-form processors. These proposed decoders have an order of magnitude lower latency and memory footprint compared to state-of-the-art decoders, while maintaining comparable throughput. In addition, we present strategies and results for implementing polar decoders on graphical processing units. Finally, we show that the energy efficiency of the proposed decoders is comparable to state-of-the-art software polar decoders.
\end{abstract}


\section{Introduction}
\label{sec:introduction}
\IEEEPARstart{I}{n} software-defined radio (SDR) applications, researchers and engineers have yet to fully harness the error-correction capability of modern codes due to their high computational complexity. Many are still using classical codes \cite{Tan2011,Demel2015} as implementing low-latency high-throughput---exceeding 10 Mbps of information throughput---software decoders for tur\-bo or low-density parity-check (LDPC) codes is very challenging. The irregular data access patterns featured in decoders of modern error-correction codes make efficient use of single-instruction multiple-data (SIMD) extensions pres\-ent on today's central processing units (CPUs) difficult. To overcome this difficulty and still achieve a good throughput, software decoders resorting to inter-frame parallelism (decoding multiple independent frames at the same time) are often proposed \cite{Xianjun2013,Wang2013,LeGal_ESL_2014}. Inter-frame parallelism comes at the cost of higher latency, as many frames have to be buffered before decoding can be started. Even with a split layer approach to LDPC decoding where intra-frame parallelism can be applied, the latency remains high at multiple milliseconds on a recent desktop processor \cite{Han2013}. This work presents software polar decoders that enable SDR systems to utilize powerful \emph{and} fast error-correction.

Polar codes provably achieve the symmetric capacity of memoryless channels \cite{Arikan2009}. Moreover they are well suited for software implementation, due to regular memory access patterns, on both x86 and embedded processors \cite{Giard2014,LeGal_SIPS_2014,LeGal_TSP_2014}.
To achieve higher throughput and lower latency on processors, software polar decoders can also exploit SIMD vector extensions present on today's CPUs. Vectorization can be performed intra-frame \cite{Giard2014} or inter-frame \cite{LeGal_SIPS_2014,LeGal_TSP_2014}, with the former having lower decoding latency as it does not require multiple frames to start decoding.

In this work, we explore intra-frame vectorized polar decoders. We propose architectures and optimization strategies that lead to the implementation of high-performance software polar decoders tailored to different processor architectures with decoding latency of 26 $\mu$s for a (32768, 29492) polar code, a significant performance improvement compared to that of our previous work \cite{Giard2014}.
We start Section~\ref{sec:background} with a review of the construction and decoding of polar codes. We then present two different software decoder architectures with varying degrees of specialization in Section~\ref{sec:compiler}. Implementation and results on an embedded processor are discussed in Section~\ref{sec:arm}. We also adapt the decoder to suit graphical processing units (GPUs), an interesting target for applications where many hundreds of frames have to be decoded simultaneously, and present the results in Section~\ref{sec:gpu}. Finally, Section~\ref{sec:energy} compares the energy consumption of the different decoders and Section~\ref{sec:conclusion} concludes the paper.

This paper builds upon the work published in \cite{Giard2014} and \cite{Sarkis_GLOBALSIP_2014}. It provides additional details on the approach as well as more experimental results for modern desktop processors. Both floating- and fixed-point implementations for the final desktop CPU version---the unrolled decoder---were further optimized leading to an information throughput of up to 1.4 Gbps. It also adds results for the adaptation of our strategies to an embedded processor leading to a throughput and latency of up to 2.25 and 36 times better, respectively, compared to that of the state-of-the-art software implementation. Compared to the state of the art, both the desktop and embedded processor implementations are shown to have one to two orders of magnitude smaller memory footprint. Lastly, strategies and results for implementing polar decoders on a graphical processing unit (GPU) are presented for the first time.

\section{Polar Codes}
\label{sec:background}

\subsection{Construction of Polar Codes}
\label{sec:polar-codes:construction}
\begin{figure}[t]
  \centering
  \subfloat[$N=2$]{\label{fig:pc2}\newcommand{\ubit}[1]{$u_{#1}$}
\newcommand{\fbit}[1]{\color{gray}$u_{#1}$}
\begin{tikzpicture}

\usetikzlibrary{shapes,positioning,arrows,decorations.markings,fit}

\definecolor{varnode_fill}{RGB}{0,0,0}
\definecolor{chknode_fill}{RGB}{255,255,255}

\tikzset{
  chknode/.style={draw,fill=chknode_fill,circle,minimum size=0.3cm, inner sep=0},
  varnode/.style={draw,fill=varnode_fill,circle,minimum size=0.1cm, inner sep=0},
  channel/.style={draw,fill=white,rectangle},
  sep/.style={rectangle,minimum width=0.3cm, inner sep=0},
  bit/.style={circle, inner sep = 0}
}

\matrix[row sep=1mm, column sep=1mm] {
 	\node[bit] (n0s0) {\ubit{0}}; & \node[sep] {}; & \node[chknode] (n0s1) {$+$}; &  \node[sep] {}; & \node[channel] (n0s2) {$W$}; & \node[sep] {}; & \node[bit] (n0s3) {$y_0$};\\
 	\node[bit] (n1s0) {\ubit{1}}; & \node[sep] {}; & \node[varnode] (n1s1) {}; &  \node[sep] {}; & \node[channel] (n1s2) {$W$}; & \node[sep] {}; & \node[bit] (n1s3) {$y_1$};\\
};

\path[-] (n0s0) edge (n0s1);
\path[-] (n0s1) edge (n0s2);
\path[-] (n0s2) edge (n0s3);

\path[-] (n1s0) edge (n1s1);
\path[-] (n1s1) edge (n1s2);
\path[-] (n1s2) edge (n1s3);

\path[-] (n1s1) edge (n0s1);

\end{tikzpicture}}
  \subfloat[$N=4$]{\label{fig:pc4}\newcommand{\ubit}[1]{$u_{#1}$}
\newcommand{\fbit}[1]{\color{gray}$u_{#1}$}
\begin{tikzpicture}

\usetikzlibrary{shapes,positioning,arrows,decorations.markings,fit}

\definecolor{varnode_fill}{RGB}{0,0,0}
\definecolor{chknode_fill}{RGB}{255,255,255}

\tikzset{
  chknode/.style={draw,fill=chknode_fill,circle,minimum size=0.3cm, inner sep=0},
  varnode/.style={draw,fill=varnode_fill,circle,minimum size=0.1cm, inner sep=0},
  channel/.style={draw,fill=white,rectangle},
  sep/.style={rectangle,minimum width=0.31cm, inner sep=0},
  bit/.style={circle, inner sep = 0}
}

\tikzset{blue dotted/.style={draw=blue!50!white, line width=1pt,
    dash pattern=on 4pt off 4pt,
    inner sep=0.5mm, rectangle, rounded corners}};

\tikzset{blue dotted tight/.style={draw=blue!50!white, line width=1pt,
    dash pattern=on 4pt off 4pt,
    inner sep=0mm, rectangle, rounded corners}};

\matrix[row sep=1mm, column sep=1mm] {
  \node[bit] (n0u) {\ubit{0}}; & \node[sep] {}; & \node[chknode] (n0s1) {$+$}; & \node[sep,label=$v_0$] {}; && \node[chknode] (n0s2) {$+$}; &  \node[sep,label={$x_0$}] {}; & \node[channel] (n0c) {$W$}; & \node[sep] {}; & \node[bit] (n0y) {$y_0$};\\
  \node[bit] (n1u) {\ubit{1}}; & \node[sep] {}; & \node[varnode] (n1s1) {};    & \node[sep,label=$v_1$] {}; &  \node[chknode] (n1s2) {$+$}; && \node[sep,label={$x_1$}] {}; & \node[channel] (n1c) {$W$}; & \node[sep] {}; & \node[bit] (n1y) {$y_1$};\\
  \node[bit] (n2u) {\ubit{2}}; & \node[sep] {}; & \node[chknode] (n2s1) {$+$}; & \node[sep,label=$v_2$] {}; && \node[varnode] (n2s2) {};    &  \node[sep,label={$x_2$}] {}; & \node[channel] (n2c) {$W$}; & \node[sep] {}; & \node[bit] (n2y) {$y_2$};\\
  \node[bit] (n3u) {\ubit{3}}; & \node[sep] {}; & \node[varnode] (n3s1) {};    & \node[sep,label=$v_3$] {}; &  \node[varnode] (n3s2) {};    && \node[sep,label={$x_3$}] {}; & \node[channel] (n3c) {$W$}; & \node[sep] {}; & \node[bit] (n3y) {$y_3$};\\
};

\path[-] (n0u) edge (n0s1);
\path[-] (n0s1) edge (n0s2);
\path[-] (n0s2) edge (n0c);
\path[-] (n0c) edge (n0y);

\path[-] (n1u) edge (n1s1);
\path[-] (n1s1) edge (n1s2);
\path[-] (n1s2) edge (n1c);
\path[-] (n1c) edge (n1y);

\path[-] (n1s1) edge (n0s1);

\path[-] (n2u) edge (n2s1);
\path[-] (n2s1) edge (n2s2);
\path[-] (n2s2) edge (n2c);
\path[-] (n2c) edge (n2y);

\path[-] (n3u) edge (n3s1);
\path[-] (n3s1) edge (n3s2);
\path[-] (n3s2) edge (n3c);
\path[-] (n3c) edge (n3y);

\path[-] (n3s1) edge (n2s1);

\path[-] (n3s2) edge (n1s2);
\path[-] (n2s2) edge (n0s2);

\end{tikzpicture}}
  \caption{Construction of polar codes of lengths 2 and 4}
\end{figure}
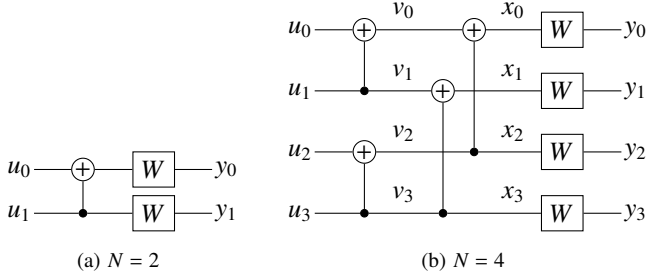

Polar codes exploit the channel polarization phenomenon to achieve the symmetric capacity of a memoryless channel as the code length increases ($N \to \infty$). A polarizing construction where $N = 2$ is shown in Fig.~\ref{fig:pc2}. The probability of correctly estimating bit $u_1$ increases compared to when the bits are transmitted without any transformation over the channel $W$. Meanwhile, the probability of correctly estimating bit $u_0$ decreases. The polarizing transformation can be combined recursively to create longer codes, as shown in Fig.~\ref{fig:pc4} for $N = 4$. As the $N \to \infty$, the probability of successfully estimating each bit approaches either 1 (perfectly reliable) or 0.5 (completely unreliable), and the proportion of reliable bits approaches the symmetric capacity of $W$ \cite{Arikan2009}.

To construct an ($N$, $k$) polar code, the $N - k$ least reliable bits, called the frozen bits, are set to zero and the remaining $k$ bits are used to carry information. The frozen bits of an (8, 5) polar code are indicated in gray in Fig.~\ref{fig:sc-graph}. The locations of the information and frozen bits are based on the type and conditions of $W$. In this work we use polar codes constructed according to \cite{Tal2011a}.
The generator matrix, $G_N$, for a polar code of length $N$ can be specified recursively so that $G_N  = F_N = F_2^{\otimes \log_2 N}$, where $F_2 = \left[ \begin{smallmatrix} 1 & 0 \\ 1 & 1\end{smallmatrix} \right]$ and $^{\otimes}$ is the Kronecker power. For example, for $N=4$, $G_N$ is
\[
G_4 = F_2^{\otimes 2} =
\left[ \begin{array}{c|c}
  F_2 & 0 \\ \hline
  F_2 & F_2 \\
\end{array} \right]
=
\begin{bmatrix}
  1 & 0 & 0 & 0\\
  1 & 1 & 0 & 0\\
  1 & 0 & 1 & 0\\
  1 & 1 & 1 & 1\\
\end{bmatrix}~.
\]

In \cite{Arikan2009}, bit-reversed indexing is used, which changes the generator matrix by multiplying it with a bit-reversal operator $B$, so that $G = B F$. In this work, natural indexing is used as it yields more efficient software decoders \cite{Giard2014}.

\subsection{Tree Representation of Polar Codes}
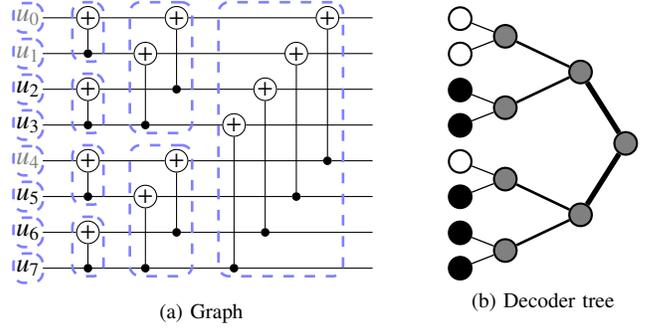
\begin{figure}[t]
  \centering
  \subfloat[Graph]{\label{fig:sc-graph}\newcommand{\ubit}[1]{$u_{#1}$}
\newcommand{\fbit}[1]{\color{gray}$u_{#1}$}
\begin{tikzpicture}[baseline=(s37.center)]

\usetikzlibrary{shapes,positioning,arrows,decorations.markings,fit}

\definecolor{varnode_fill}{RGB}{0,0,0}
\definecolor{chknode_fill}{RGB}{255,255,255}

\tikzset{
  chknode/.style={draw,fill=chknode_fill,circle,minimum size=0.3cm, inner sep=0},
  varnode/.style={draw,fill=varnode_fill,circle,minimum size=0.1cm, inner sep=0},
  sep/.style={rectangle,minimum width=0.25cm, inner sep=0},
  bit/.style={circle, inner sep = 0}
}

\tikzset{blue dotted/.style={draw=blue!50!white, line width=1pt,
    dash pattern=on 4pt off 4pt,
    inner sep=0.5mm, rectangle, rounded corners}};

\tikzset{blue dotted tight/.style={draw=blue!50!white, line width=1pt,
    dash pattern=on 4pt off 4pt,
    inner sep=0mm, rectangle, rounded corners}};

\matrix[row sep=1mm, column sep=1mm] {
	\node[bit] (n0s0) {\fbit{0}}; & \node[sep] (s0) {}; & \node[chknode] (n0s1) {$+$}; & \node[sep] (s10) {}; &&	\node[chknode] (n0s2) {$+$}; &	\node[sep] (s20) {}; &&&& \node[chknode] (n0s3) {$+$}; &    \node[sep] (s30) {}; & \node[circle] (n0s4) {}; \\ 
	\node[bit] (n1s0) {\fbit{1}}; & \node[sep] (s1) {}; & \node[varnode] (n1s1) {};	   & \node[sep] (s11) {}; &	 \node[chknode] (n1s2) {$+$}; && \node[sep] (s21) {}; &&&  \node[chknode] (n1s3) {$+$}; &&   \node[sep] (s31) {}; & \node[circle] (n1s4) {}; \\ 
	\node[bit] (n2s0) {\ubit{2}}; & \node[sep] (s2) {}; & \node[chknode] (n2s1) {$+$}; & \node[sep] (s12) {}; &&	 \node[varnode] (n2s2) {};    &	 \node[sep] (s22) {}; &&   \node[chknode] (n2s3) {$+$}; &&&  \node[sep] (s32) {}; & \node[circle] (n2s4) {}; \\ 
	\node[bit] (n3s0) {\ubit{3}}; & \node[sep] (s3) {}; & \node[varnode] (n3s1) {};	   & \node[sep] (s13) {}; &	 \node[varnode] (n3s2) {};    && \node[sep] (s23) {}; &	   \node[chknode] (n3s3) {$+$}; &&&& \node[sep] (s33) {}; & \node[circle] (n3s4) {}; \\ 
	\node[bit] (n4s0) {\fbit{4}}; & \node[sep] (s4) {}; & \node[chknode] (n4s1) {$+$}; & \node[sep] (s14) {}; &&	 \node[chknode] (n4s2) {$+$}; &	 \node[sep] (s24) {}; &&&& \node[varnode] (n4s3) {};	&    \node[sep] (s34) {}; & \node[circle] (n4s4) {}; \\ 
	\node[bit] (n5s0) {\ubit{5}}; & \node[sep] (s5) {}; & \node[varnode] (n5s1) {};	   & \node[sep] (s15) {}; &	 \node[chknode] (n5s2) {$+$}; && \node[sep] (s25) {}; &&&  \node[varnode] (n5s3) {};	&&   \node[sep] (s35) {}; & \node[circle] (n5s4) {}; \\ 
	\node[bit] (n6s0) {\ubit{6}}; & \node[sep] (s6) {}; & \node[chknode] (n6s1) {$+$}; & \node[sep] (s16) {}; &&	 \node[varnode] (n6s2) {};    &	 \node[sep] (s26) {}; &&   \node[varnode] (n6s3) {};	&&&  \node[sep] (s36) {}; & \node[circle] (n6s4) {}; \\ 
	\node[bit] (n7s0) {\ubit{7}}; & \node[sep] (s7) {}; & \node[varnode] (n7s1) {};	   & \node[sep] (s17) {}; &	 \node[varnode] (n7s2) {};    && \node[sep] (s27) {}; &	   \node[varnode] (n7s3) {};	&&&& \node[sep] (s37) {}; & \node[circle] (n7s4) {}; \\ 
};
\path[-] (n0s0) edge (n0s1) (n0s1) edge (n0s2) (n0s2) edge (n0s3) (n0s3) edge (n0s4);
\path[-] (n1s0) edge (n1s1) (n1s1) edge (n1s2) (n1s2) edge (n1s3) (n1s3) edge (n1s4);
\path[-] (n2s0) edge (n2s1) (n2s1) edge (n2s2) (n2s2) edge (n2s3) (n2s3) edge (n2s4);
\path[-] (n3s0) edge (n3s1) (n3s1) edge (n3s2) (n3s2) edge (n3s3) (n3s3) edge (n3s4);
\path[-] (n4s0) edge (n4s1) (n4s1) edge (n4s2) (n4s2) edge (n4s3) (n4s3) edge (n4s4);
\path[-] (n5s0) edge (n5s1) (n5s1) edge (n5s2) (n5s2) edge (n5s3) (n5s3) edge (n5s4);
\path[-] (n6s0) edge (n6s1) (n6s1) edge (n6s2) (n6s2) edge (n6s3) (n6s3) edge (n6s4);
\path[-] (n7s0) edge (n7s1) (n7s1) edge (n7s2) (n7s2) edge (n7s3) (n7s3) edge (n7s4);

\path[-] (n0s1) edge (n1s1);
\path[-] (n2s1) edge (n3s1);
\path[-] (n4s1) edge (n5s1);
\path[-] (n6s1) edge (n7s1);

\path[-] (n0s2) edge (n2s2);
\path[-] (n1s2) edge (n3s2);
\path[-] (n4s2) edge (n6s2);
\path[-] (n5s2) edge (n7s2);

\path[-] (n0s3) edge (n4s3);
\path[-] (n1s3) edge (n5s3);
\path[-] (n2s3) edge (n6s3);
\path[-] (n3s3) edge (n7s3);

\node (g_n0s0) [blue dotted tight, fit = (n0s0)] {};
\node (g_n1s0) [blue dotted tight, fit = (n1s0)] {};
\node (g_n2s0) [blue dotted tight, fit = (n2s0)] {};
\node (g_n3s0) [blue dotted tight, fit = (n3s0)] {};
\node (g_n4s0) [blue dotted tight, fit = (n4s0)] {};
\node (g_n5s0) [blue dotted tight, fit = (n5s0)] {};
\node (g_n6s0) [blue dotted tight, fit = (n6s0)] {};
\node (g_n7s0) [blue dotted tight, fit = (n7s0)] {};

\node (g_n0s1) [blue dotted, fit = (n0s1) (n1s1)] {};
\node (g_n1s1) [blue dotted, fit = (n2s1) (n3s1)] {};
\node (g_n2s1) [blue dotted, fit = (n4s1) (n5s1)] {};
\node (g_n3s1) [blue dotted, fit = (n6s1) (n7s1)] {};

\node (g_n0s2) [blue dotted, fit = (n0s2) (n1s2) (n2s2) (n3s2)] {};
\node (g_n1s2) [blue dotted, fit = (n4s2) (n5s2) (n6s2) (n7s2)] {};

\node (g_n0s3) [blue dotted, fit = (n0s3) (n1s3) (n2s3) (n3s3) (n4s3) (n5s3) (n6s3) (n7s3)] {};

\end{tikzpicture}}
  \quad\hspace{-5pt}
  \subfloat[Decoder tree]{\label{fig:sc-tree-1} \begin{tikzpicture}[baseline = (0_7.center),
        level/.style={level distance = 6mm},
        level 1/.style={sibling distance=19mm, edge from parent/.style={draw,black,line width=2pt}},
        level 2/.style={level distance=10mm, sibling distance=9.5mm, edge from parent/.style={draw,black,line width=1pt}},
        level 3/.style={sibling distance=4.7mm, edge from parent/.style={draw,black,line width=0.5pt}},
        ]

\tikzset{
frozen/.style={thick,draw=black,fill=white,minimum size=3mm,circle, inner sep=0},
fullspace/.style={thick,draw=black,fill=black,minimum size=3mm,circle, inner sep = 0},
mixed/.style={thick,draw=black,fill=gray,minimum size=3mm,circle, inner sep = 0},
ml_mixed/.style={thick,draw=black,fill=blue,minimum size=3mm,circle, inner sep = 0}
}

\node[mixed] (p){} [grow=left]
	child {node[mixed] (2_0){}
		child {node[mixed] (1_0){}
			child {node[frozen] (a0_0){}
			}
			child {node[frozen] (a0_1){}
			}
		}
		child {node[mixed] (1_2){}
			child {node[fullspace] (0_2){}
			}
			child {node[fullspace] (0_3){}
			}
		}
	}
	child {node[mixed] (v){}
		child {node[mixed] (cl){}
			child {node[frozen] (0_4){}
			}
			child {node[fullspace] (0_5){}
			}
		}
		child {node[mixed] (cr){}
			child {node[fullspace] (0_6){}
			}
			child {node[fullspace] (0_7){}
			}
		}
	}
;

\end{tikzpicture}}
  \caption{The graph and tree representation of an $(8,5)$ polar code.}
  \label{fig:graph-to-tree}
\end{figure}

A polar code of length $N$ is the concatenation of two constituent polar codes of length $\nicefrac{N}{2}$ \cite{Arikan2009}. Therefore, binary trees are a natural representation of polar codes \cite{Alamdar-Yazdi2011}. Fig.~\ref{fig:graph-to-tree} illustrates the tree representation of an (8, 5) polar code. In Fig.~\ref{fig:sc-graph}, the frozen bits are labeled in gray while the information bits are in black. The corresponding tree, shown in Fig.~\ref{fig:sc-tree-1}, uses white and black leaf nodes to denote these bits, respectively. The gray nodes of Fig.~\ref{fig:sc-tree-1} correspond to concatenation operations shown in Fig.~\ref{fig:sc-graph}.

\subsection{Successive-Cancellation Decoding}
In successive-cancellation (SC) decoding, the decoder tree is traversed depth first, selecting left edges before backtracking to right ones, until the size-1 frozen and information leaf nodes. The messages passed to child nodes are log-likelihood ratios (LLRs); while those passed to parents are bit estimates. These messages are denoted $\alpha$ and $\beta$, respectively. Messages to a left child $l$ are calculated by the $f$ operation using the min-sum algorithm:
\begin{equation}
\label{eq:f}
\begin{split}
\alpha_l[i] & = f( \alpha_v[i], \alpha_v[i + \nicefrac{N_v}{2}] )\\
& = \sgn{\alpha_v[i]}\sgn{\alpha_v[i + \nicefrac{N_v}{2}]} \min(|\alpha_v[i]|, |\alpha_v[i + \nicefrac{N_v}{2}]|),
\end{split}
\end{equation}
where $N_v$ is the size of the corresponding constituent code and $\alpha_v$ the LLR input to the node.

Messages to a right child are calculated using the $g$ operation
\begin{equation}
\label{eq:g}
\begin{split}
\alpha_r[i] & = g( \alpha_v[i], \alpha_v[i + \nicefrac{N_v}{2}], \beta_l[i] )\\
& = \begin{cases}
\alpha_v[i + \nicefrac{N_v}{2}] + \alpha_v[i]\text{,} & \text{when } \beta_l[i] = 0;\\
\alpha_v[i + \nicefrac{N_v}{2}] - \alpha_v[i]\text{,} & \text{otherwise},
\end{cases}
\end{split}
\end{equation}
where $\beta_l$ is the bit estimate from the left child.

Bit estimates at the leaf nodes are set to zero for frozen bits and are calculated by performing threshold detection for information ones. After a node has the bit estimates from both its children, they are combined to generate the node's estimate that is passed to its parent
\begin{equation}
\label{eq:combine}
\beta_v[i] = \begin{cases}
\beta_l[i] \oplus \beta_r[i]\text{,} & \text{when } i < \nicefrac{N_v}{2};\\
\beta_r[i - \nicefrac{N_v}{2}]\text{,} & \text{otherwise,}
\end{cases}
\end{equation}
where $\oplus$ is modulo-2 addition (XOR).

\subsection{Simplified Successive-Cancellation Decoding}
\begin{figure}[t]
  \centering
  \subfloat[SC]{\label{fig:sc-tree-2}\begin{tikzpicture}[baseline = (0_7.center),
        level/.style={level distance = 6mm},
        level 1/.style={sibling distance=19mm, edge from parent/.style={draw,black,line width=2pt}},
        level 2/.style={level distance=10mm, sibling distance=9.5mm, edge from parent/.style={draw,black,line width=1pt}},
        level 3/.style={sibling distance=4.7mm, edge from parent/.style={draw,black,line width=0.5pt}},
        ]

\tikzset{
frozen/.style={thick,draw=black,fill=white,minimum size=3mm,circle, inner sep=0},
fullspace/.style={thick,draw=black,fill=black,minimum size=3mm,circle, inner sep = 0},
mixed/.style={thick,draw=black,fill=gray,minimum size=3mm,circle, inner sep = 0},
ml_mixed/.style={thick,draw=black,fill=blue,minimum size=3mm,circle, inner sep = 0}
}

\node[mixed] (p){} [grow=left]
	child {node[mixed] (2_0){}
		child {node[mixed] (1_0){}
			child {node[frozen] (a0_0){}
			}
			child {node[frozen] (a0_1){}
			}
		}
		child {node[mixed] (1_2){}
			child {node[fullspace] (0_2){}
			}
			child {node[fullspace] (0_3){}
			}
		}
	}
	child {node[mixed] (v){$v$}
		child {node[mixed, label={[label distance = -1mm]left}] (cl){}
			child {node[frozen] (0_4){}
			}
			child {node[fullspace] (0_5){}
			}
		}
		child {node[mixed, label={[label distance = -1mm]below:right}] (cr){}
			child {node[fullspace] (0_6){}
			}
			child {node[fullspace] (0_7){}
			}
		}
	}
;

\draw[<-] ($(v.north east) - (1mm, -1mm)$) -- node[above left=-1.5mm] {\footnotesize $\alpha_v$} ($(p.south west) - (1mm, 0mm)$);
\draw[<-] ($(p.south west) + (1mm, -1mm)$) -- node[below right=-2mm] {\footnotesize $\beta_v$} ($(v.north east) + (1mm, 0mm)$);

\draw[<-] ($(v.north west) + (-1.6mm, -1mm)$) -- node[left=1mm] {\footnotesize $\beta_l$} ($(cl.south east) + (0mm, -0.6mm)$);
\draw[<-] ($(cl.south east) - (-1.6mm, -1mm)$) -- node[above right=-2mm] {\footnotesize$\alpha_l$} ($(v.north west) - (0mm, -0.6mm)$);

\draw[<-] ($(v.south west) + (0mm, -0.6mm)$) -- node[below right=-2mm] {\footnotesize $\beta_r$} ($(cr.north east) + (1.6mm, -1mm)$);
\draw[<-] ($(cr.north east) - (0mm, -0.6mm)$) -- node[left=1mm] {\footnotesize $\alpha_r$} ($(v.south west) - (1.6mm, -1mm)$);

\end{tikzpicture}}
  \quad
  \subfloat[SSC]{\label{fig:ssc-tree} \begin{tikzpicture}[baseline=(base.center),
        level/.style={level distance = 6mm},
        level 1/.style={sibling distance=19mm, edge from parent/.style={draw,black,line width=2pt}},
        level 2/.style={sibling distance=9mm, edge from parent/.style={draw,black,line width=1pt}},
        level 3/.style={sibling distance=4mm, edge from parent/.style={draw,black,line width=0.5pt}},
        ]

\tikzset{
frozen/.style={thick,draw=black,fill=white,minimum size=3mm,circle, inner sep=0},
fullspace/.style={thick,draw=black,fill=black,minimum size=3mm,circle, inner sep = 0},
mixed/.style={thick,draw=black,fill=gray,minimum size=3mm,circle, inner sep = 0},
ml_mixed/.style={thick,draw=black,fill=blue,minimum size=3mm,circle, inner sep = 0}
}

\node[mixed] (3_0){} [grow=left]
	child {node[mixed] (2_0){}
  	child {node[frozen] (0_4){}
    }
		child {node[fullspace] (1_2){}
		}
	}
	child {node[mixed] (2_1){}
		child {node[mixed] (1_2){}
			child {node[frozen] (0_4){}
			}
			child {node[fullspace] (0_5){}
			}
		}
		child {node[fullspace] (1_3){}
		}
	}
;

\node [circle,below= 0.27mm of 1_3.base] (base) {};

\end{tikzpicture}}
  \quad
  \subfloat[Fast-SSC]{\label{fig:fast-ssc-tree} \definecolor{deepgreen}{RGB}{8, 130, 25}

\begin{tikzpicture}[baseline=(base),
        level/.style={level distance = 6mm},
        level 1/.style={sibling distance=19mm, edge from parent/.style={draw,black,line width=2pt}},
        level 2/.style={sibling distance=9mm, edge from parent/.style={draw,black,line width=1pt}},
        level 3/.style={sibling distance=4mm, edge from parent/.style={draw,black,line width=0.5pt}},
        ]

\tikzset{
frozen/.style={thick,draw=black,fill=white,minimum size=3mm,circle, inner sep=0},
hidden/.style={draw=white,fill=white,minimum size=3mm,circle, inner sep=0},
fullspace/.style={thick,draw=black,fill=black,minimum size=3mm,circle, inner sep = 0},
mixed/.style={thick,draw=black,fill=gray,minimum size=3mm,circle, inner sep = 0},
rep_mixed/.style={thick,draw=black,pattern=north west lines,pattern color=deepgreen,minimum size=3mm,circle, inner sep = 0},
spc_mixed/.style={thick,draw=black,pattern=crosshatch,pattern color=orange,minimum size=3mm,circle, inner sep = 0},
repspc/.style={thick,draw=black,pattern=vertical lines,pattern color=blue,minimum size=3mm,circle, inner sep = 0}
}

\node[mixed] (3_0){} [grow=left]
	child {node[shape=circle split, circle split part fill={white,gray},
    scale=0.625, draw=black, line width=0.275mm] (2_0){}
    child {node[fullspace,yshift=-4mm] (1_2){}
    }
	}
	child {node[spc_mixed,label={[label distance=0cm]-2:{\footnotesize SPC}}] (2_1){}
	}
;
	
\node [circle, below= 5.27mm of 2_1.base] (base) {};
\node[circle, left= 4mm of 3_0] (pad1) {};
\end{tikzpicture}}
  \caption{Decoder trees corresponding to the SC, SSC and Fast-SSC decoding algorithms.}
  \label{fig:trees}
\end{figure}
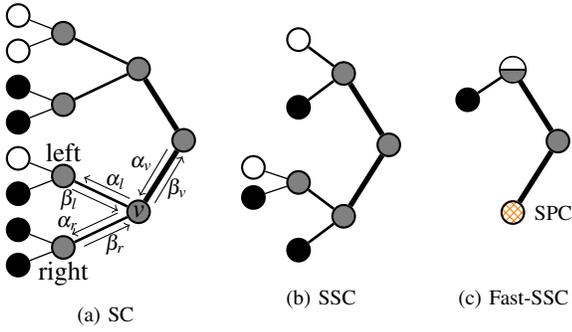

Instead of traversing a sub-tree whose leaves all correspond to frozen or information bits, simplified successive-cancella\-tion (SSC) applies a decision rule immediately \cite{Alamdar-Yazdi2011}. For fro\-zen sub-trees, the output is set to the zero vector; while for information sub-tree the maximum-likelihood (ML) output is obtained by performing element-wise threshold detection on the soft-information input vector, $\alpha_v$. This shrinks the decoder, reducing the number of calculations and increasing decoding speed. The SC and SSC pruned tree corresponding to an $(8, 5)$ polar code are shown in Fig.~\ref{fig:sc-tree-2} and Fig.~\ref{fig:ssc-tree}, respectively.

\subsection{The Fast-SSC Decoding Algorithm}
The Fast-SSC decoding algorithm further prunes the deco-der tree by applying low-complexity decoding rules when encountering certain types of constituent codes. These special cases are:

\textit{Repetition codes: }%
are constituent codes where only the last bit is an information bit.
These codes are efficiently decoded by calculating the sum of the input LLRs and using threshold detection to determine the result that is then replicated to form the estimated bits :
\[
\beta_v[i] = \begin{cases}
0, & \text{when } \left(\sum_{i=0}^{N_v-1}{\mvec{\alpha}_v[i]}\right) \geq 0;\\
1, & \text{otherwise,}
\end{cases}
\]
where $N_v$ is the number of leaf nodes.

\textit{Single-parity-check (SPC) codes: }%
are constituent codes where only the first bit is frozen. The corresponding node is indicated by the cross-hatched orange pattern in Fig.~\ref{fig:fast-ssc-tree}. The first step in decoding these codes is to calculate the hard decision of each LLR
\begin{equation}\label{eq:info}
\mvec{\beta}_v[i] = \begin{cases}
0, & \text{when } \mvec{\alpha}_v[i] \geq 0;\\
1, & \text{otherwise,}
\end{cases}
\end{equation}
and then calculating the parity of these decisions
\begin{equation}
\label{eq:spc:parity}
\text{parity} = \bigoplus_{i = 0}^{N_v-1} \mvec{\beta}_v[i]\nonumber.
\end{equation}
If the parity constraint is unsatisfied, the estimate of the bit with the smallest LLR magnitude is flipped:
\begin{equation}
\label{eq:spc:beta}
\mvec{\beta}_v[i] = \mvec{\beta}_v[i] \oplus \text{parity}, \text{where } i = \underset{i}{\arg\, \min}(|\mvec{\alpha}_v[i]|).\nonumber
\end{equation}

\textit{Repetition-SPC codes: }%
are codes whose left constituent code is a repetition code and the right an SPC one. They can be speculatively decoded in hardware by simultaneously decoding the repetition code and two instances of the SPC code: one assuming the output of the repetition code is all 0's and the other all 1's. The correct result is selected once the output of the repetition code is available. This speculative decoding also provides speed gains in software.\\

Fig.~\ref{fig:fast-ssc-tree} shows the tree corresponding to a Fast-SSC decoder is will be described more thoroughly in Section~\ref{sec:impl:unrolled}. Other types of operations are introduced in the Fast-SSC algorithm, we refer the reader to \cite{Sarkis2014} for more details.

\section{Implementation on x86 Processors}
\label{sec:compiler}
In this section we present two different versions of the decoder in terms of increasing design specialization for software; whereas the first version---the instruction-based deco\-der---takes advantage of the processor architecture it remains configurable at run time and the second one---the unrolled deco\-der---presents a fully unrolled, branchless decoder fully exploiting SIMD vectorization. In the second version of the decoder, compile-time optimization plays a significant role in the performance improvements. Performance is evaluated for both the instruction-based and unrolled decoders.

It should be noted that, contrary to what is common in hardware implementations e.g. \cite{Leroux2013,Sarkis2014}, natural indexing is used for all software decoder implementations. While bit-reversed indexing is well-suited for hardware decoders, SIMD instructions operate on independent vectors, not adjacent values within a vector. Using bit-reverse indexing would have mandated data shuffling operations before any vectorized operation is performed.

Both versions, instruction-based decoders and unrolled decoders, use the following functions from the Fast-SSC algorithm \cite{Sarkis2014}: F, G, G\_0R, Combine, Combine\_0R, Repetition, 0SPC, RSPC, RepSPC and P\_01. An Info function implementing eq.~\eqref{eq:info} is also added.

\subsubsection*{Methodology for the Experimental Results}
We discuss throughput in information bits per second as well as latency. Our software was compiled using the C++ compiler from GCC 4.9 using the flags ``\texttt{-march=native -funroll-loops -Ofast}''. Additionally, auto-vectori\-za\-ti\-on is always kept enabled. The decoders are inserted in a digital communication chain to measure their speed and to ensure that optimizations, including those introduced by\\\texttt{-Ofast}, do not affect error-correction performance. In the simulations, we use binary phase shift keying (BPSK) over an AWGN channel with random codewords.

The throughput is calculated using the time required to decode a frame averaged over 10 runs of $\num[group-separator={,}]{50000}$ and $\num[group-separator={,}]{10000}$ frames each for the $N=2048$ and the $N>2048$ codes, respectively. The time required to decode a frame, or latency, also includes the time required to copy a frame to decoder memory and copy back the estimated codeword. Time is measured using the high precision clock provided by the Boost Chrono library.

In this work we focus on decoders running on one processor core only since the targeted application is SDR. Typically, an SDR system cannot afford to dedicate more than a single core to error-correction as it has to perform other functions simultaneously. For example, in SDR implementations of long term evolution (LTE) receivers, the orthogonal frequency-division multiplexing (OFDM) demodulation alone is approximately an order of magnitude more computationally demanding than the error-correction decoder \cite{Tan2011,Bang2014,Demel2015}.

\subsection{Instruction-based Decoder}
\label{sec:impl:ni}
The Fast-SSC decoder implemented on a field-programma\-ble gate array (FPGA) in \cite{Sarkis2014} closely resembles a CPU with wide SIMD vector units and wide data buses. Therefore, it was natural to use the same design for a software decoder, leveraging SIMD instructions. This section describes how the algorithm was adapted for a software implementation. As fixed-point arithmetic can be used, the effect of quantization is shown.

\subsubsection{Using Fixed-Point Numbers}
\begin{figure}[t]
  \centering
\begin{tikzpicture}

  \pgfplotsset{
    grid style = {
      dash pattern = on 0.05mm off 1mm,
      line cap = round,
      black,
      line width = 0.5pt
    },
    label style = {font=\fontsize{9pt}{7.2}\selectfont},
    tick label style = {font=\fontsize{7pt}{7.2}\selectfont}
  }

  \draw (1.4,4.4) node[above]{\fontsize{9pt}{7.2}\selectfont$(2048,1024)$};

  \begin{semilogyaxis}[%
    xlabel=$E_b/N_0$ (dB),xtick={2,2.5,3,3.5,4,4.5},%
    xmin=1.75,%
    xlabel style={yshift=0.6em},%
    ylabel=Error rate, ylabel style={yshift=-1.2em},%
    ytick={1e-2,1e-4,1e-6,1e-8,1e-10,1e-12},%
    width=0.5\columnwidth, height=6cm, grid=major,%
    legend style={
      anchor={center},
      cells={anchor=west},
      column sep=2mm,
      font=\footnotesize,
    },
    legend columns=2,
    legend to name=perf-legend,
    mark size=3.0pt]

    \addplot[color=black] table[x=snr_db,y=BER]{2k5.s0.708.float.txt};
    \addlegendentry{BER Float}

    \addplot[color=black,densely dashed] table[x=snr_db,y=FER]{2k5.s0.708.float.txt};
    \addlegendentry{FER Float}

    \addplot[color=red,mark=star] table[x=snr_db,y=BER]{2k5.s0.708.int8.txt};
    \addlegendentry{BER 8-bit}
    \addplot[color=red,mark=star,densely dashed] table[x=snr_db,y=FER]{2k5.s0.708.int8.txt};
    \addlegendentry{FER 8-bit}

 \end{semilogyaxis}
\end{tikzpicture}
\begin{tikzpicture}

\pgfplotsset{
  grid style = {
    dash pattern = on 0.05mm off 1mm,
    line cap = round,
    black,
    line width = 0.5pt
  },
  label style = {font=\fontsize{9pt}{7.2}\selectfont},
  tick label style = {font=\fontsize{7pt}{7.2}\selectfont}
}

\draw (1.4,4.4) node[above]{\fontsize{9pt}{7.2}\selectfont$(32768,27568)$};

\begin{semilogyaxis}[%
    xlabel=$E_b/N_0$ (dB),xtick={3.75,4.00,4.25,4.5},%
    ylabel=Error rate, ylabel style={yshift=-1.2em},%
    xlabel style={yshift=0.6em},%
    ytick={1e-2,1e-4,1e-6,1e-8,1e-10,1e-12},%
    width=0.5\columnwidth, height=6cm, grid=major,%
    mark size=3.0pt]

    \addplot[color=black] table[x=snr_db,y=BER]{32k.0.84.float.txt};

    \addplot[color=black,densely dashed] table[x=snr_db,y=FER]{32k.0.84.float.txt};

    \addplot[color=red,mark=star] table[x=snr_db,y=BER]{32k.0.84.int8.txt};

    \addplot[color=red,mark=star,densely dashed] table[x=snr_db,y=FER]{32k.0.84.int8.txt};

  \end{semilogyaxis}
\end{tikzpicture}
\\
\ref{perf-legend}
  \caption{Effect of quantization on error-correction performance.}
  \label{fig:perf_quant}
\end{figure}
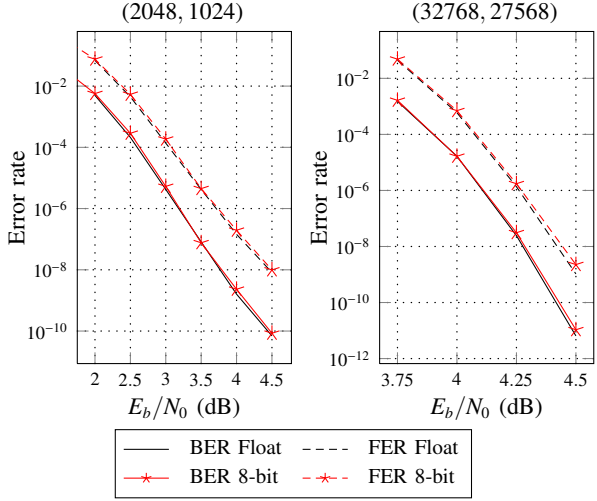

On processors, fixed-point numbers are represented with at least 8 bits. As illustrated in Fig.~\ref{fig:perf_quant}, using 8 bits of quantization for LLRs results in a negligible degradation of error-correction performance over a floating-point representation. At a frame-error rate (FER) of $10^{-8}$ the performance loss compared to a floating-point implementation is less than \\$0.025$~dB for the (32768, 27568) polar code. With custom hardware, it was shown in \cite{Sarkis2014} that 6 bits are sufficient for that polar code. It should be noted that in Fast-SSC decoding, only the G function adds to the amplitude of LLRs and it is carried out with saturating adders.

With instructions that can work on registers of packed 8-bit integers, the SIMD extensions available on most general-purpose x86 and ARM processors are a good fit to implement a polar decoder.

\subsubsection{Vectorizing the Decoding of Constituent Codes}
On x86-64 processors, the vector instructions added with SSE support logic and arithmetic operations on vectors containing either 4 single-precision floating-point numbers or 16 8-bit integers. Additionally, x86-64 processors with AVX instructions can operate on data sets of twice that size. Below are the operations benefiting the most from explicit vectorization.

$F$: the $f$ operation \eqref{eq:f} is often executed on large vectors of LLRs to prepare values for other processing nodes. The $\min()$ operation and the sign calculation and assignment are all vectorized.

$G$ and $G\_0R$: the $g$ operation is also frequently executed on large vectors. Both possibilities, the sum and the difference, of \eqref{eq:g} are calculated and are blended together with a mask to build the result. The G\_0R operation replaces the G operation when the left hand side of the tree is the all-zero vector.

\textit{Combine} and \textit{Combine\_0R}: %
the Combine operation combines two estimated bit-vectors using an XOR operation in a vectorized manner. The Combine\_0R operation is to Combine what G\_0R is to G.

\textit{SPC decoding}: %
locating the LLR with the minimum magnitude is accelerated using SIMD instructions.

\subsubsection{Data Representation}
For the decoders using floating-point numbers, the representation of $\beta$ is changed to accelerate the execution of the $g$ operation on large vectors. Thus, when floating-point LLRs are used, $\beta_l[i] \in \{+1, -1\}$ instead of $\{0, 1\}$. As a result, \eqref{eq:g} can be rewritten as
\[
g( \alpha_v[i], \alpha_v[i + \nicefrac{N_v}{2}], \beta_l[i] ) = \alpha_v[i]*\beta_l[i] + \alpha_v[i + \nicefrac{N_v}{2}].
\]
This removes the conditional assignment and turns $g()$ into a multiply-accumulate operation, which can be performed efficiently in a vectorized manner on modern CPUs. For integer LLRs, multiplications cannot be carried out on 8-bit integers. Thus, both possibilities of \eqref{eq:g} are calculated and are blended together with a mask to build the result. The Combine operation is modified accordingly for the floating-point decoder and is computed using a multiplication with $\beta_l[i] \in \{+1, -1\}$.

\subsubsection{Architecture-specific Optimizations}
The decoders take advantage of the SSSE 3, SSE 4.1 and AVX instructions when available. Notably, the \texttt{sign} and \texttt{abs} instructions from SSSE 3 and the \texttt{blendv} instruction from SSE 4.1 are used. AVX, with instructions operating on vectors of 256 bits instead of the 128 bits, is only used for the floating-point implementation since it does not support integer operations. Data was aligned to the 128 (SSE) or 256-bit (AVX) boundaries for faster accesses.

\subsubsection{Implementation Comparison}
\label{sect:cmp_v1}
Here we compare the performance of three implementations. First, a non-explicitly vectorized version using floating-point numbers. Second an explicitly vectorized version using float\-ing-point numbers. Third, the explicitly vectorized version using a fixed-point number representation. In Table~\ref{tab:impl:tp2}, they are denoted as Float, SIMD-Float and SIMD-int8 respectively.

Results for decoders using the floating-point number representation are included as the efficient implementation ma\-kes the resulting throughput high enough for some applications. The decoders ran on a single core of an Intel Core i7-4770S clocked at 3.1 GHz with Turbo disabled.

Comparing the throughput and latency of the Float and SIMD-Float implementations in Table~\ref{tab:impl:tp2} confirms the benefits of explicit vectorization in this decoder. The performance of the SIMD-Float implementation is only 21\% to 38\% slower than the SIMD-int8 implementation. This is not a surprising result considering that the SIMD-Float implementation uses the AVX instructions operating on vectors of 256 bits while the SIMD-int8 version is limited to vectors of 128 bits. Table~\ref{tab:impl:tp2} also shows that vectorized implementations have 3.6 to 5.8 times lower latency than the floating-point decoder.

\begin{table}[t]
  \centering
  \caption{Decoding polar codes with the instruction-based decoder.}
  \begin{tabular}{c c C{1.2cm} c}
    \toprule
    \multirow{2}{*}{\shortstack{Code\\$(N,k)$}} & \multirow{2}{*}{Implementation} & Info T/P & \multirow{2}{*}{\shortstack{Latency \\ ($\mu$s)}} \\
                          &                    & (Mbps) &\\
    \midrule
    $(2048,1024)$      & Float              &  20.8 &  49\\
                       & SIMD-Float         &  75.6 &  14\\
                       & SIMD-int8          & 121.7 &   8\vspace{2pt}\\
    $(2048,1707)$      & Float              &  41.5 &  41\\
                       & SIMD-Float         & 173.9 &  10\\
                       & SIMD-int8          & 209.9 &   8\vspace{2pt}\\
    $(32768, 27568)$   & Float              &  32.4 & 825\\
                       & SIMD-Float         & 124.3 & 222\\
                       & SIMD-int8          & 175.1 & 157\vspace{2pt}\\
    $(32768, 29492)$   & Float              &  40.8 & 723\\
                       & SIMD-Float         & 160.1 & 184\\
                       & SIMD-int8          & 198.6 & 149\\
    \bottomrule
  \end{tabular}%
  \label{tab:impl:tp2}
\end{table}

\subsection{Unrolled Decoder}
\label{sec:impl:unrolled}
The goal of this design is to increase vectorization and inlining and reduce branches in the resulting decoder by maximizing the information specified at compile-time. It also gets rid of the indirections that were required to get good performance out of the instruction-based decoder.

\subsubsection{Generating an Unrolled Decoder}
The polar codes decoded by the instruction-based decoders presented in Section~\ref{sec:impl:ni} can be specified at run-time. This flexibility comes at the cost of increased branches in the code due to conditionals, indirections and loops.
Creating a decoder dedicated to only one polar code enables the generation of a branchless fully-unrolled decoder. In other words, knowing in advance the dimensions of the polar code and the frozen bit locations removes the need for most of the control logic and eliminates branches there.

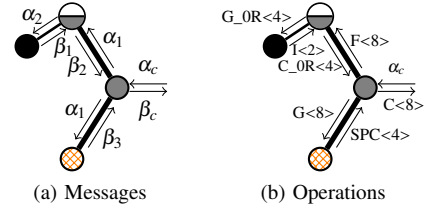
\begin{figure}[t]
  \centering
  \subfloat[Messages]{\definecolor{deepgreen}{RGB}{8, 130, 25}

\begin{tikzpicture}[baseline=(base),
        level/.style={level distance = 6mm},
        level 1/.style={sibling distance=19mm, edge from parent/.style={draw,black,line width=2pt}},
        level 2/.style={sibling distance=9mm, edge from parent/.style={draw,black,line width=1pt}},
        level 3/.style={sibling distance=4mm, edge from parent/.style={draw,black,line width=0.5pt}},
        ]

\tikzset{
frozen/.style={thick,draw=black,fill=white,minimum size=3mm,circle, inner sep=0},
fullspace/.style={thick,draw=black,fill=black,minimum size=3mm,circle, inner sep = 0},
mixed/.style={thick,draw=black,fill=gray,minimum size=3mm,circle, inner sep = 0},
rep_mixed/.style={thick,draw=black,pattern=north west lines,pattern color=deepgreen,minimum size=3mm,circle, inner sep = 0},
spc_mixed/.style={thick,draw=black,pattern=crosshatch,pattern color=orange,minimum size=3mm,circle, inner sep = 0},
repspc/.style={thick,draw=black,pattern=vertical lines,pattern color=blue,minimum size=3mm,circle, inner sep = 0}
}

\tikzset{
parallel segment/.style={
   segment distance/.store in=\segDistance,
   segment pos/.store in=\segPos,
   segment length/.store in=\segLength,
   to path={
   ($(\tikztostart)!\segPos!(\tikztotarget)!\segLength/2!(\tikztostart)!\segDistance!90:(\tikztotarget)$) -- 
   ($(\tikztostart)!\segPos!(\tikztotarget)!\segLength/2!(\tikztotarget)!\segDistance!-90:(\tikztostart)$)  \tikztonodes
   }, 
   segment pos=.5,
   segment length=4ex,
   segment distance=-1mm,
},
}

\node[mixed] (3_0){} [grow=left]
	child {node[shape=circle split, circle split part fill={white,gray},
    scale=0.625, draw=black, line width=0.275mm] (2_0){}
    child {node[fullspace,yshift=-4mm] (1_2){}
    }
	}
	child {node[spc_mixed] (2_1){}
	}
;
	
\draw[->] (3_0) to[parallel segment] node[above right=-1.5mm] {\footnotesize $\alpha_1$} (2_0);
\draw[->] (2_0) to[parallel segment, segment length=2.5ex] node[above left=-1.5mm] {\footnotesize $\alpha_2$} (1_2);
\draw[->] (1_2) to[parallel segment, segment length=2.5ex] node[below right=-1.75mm] (I2) {\footnotesize $\beta_1$} (2_0);
\draw[->] (2_0) to[parallel segment] node[below left=-1.75mm] {\footnotesize $\beta_2$} (3_0);

\draw[->] (3_0) to[parallel segment] node[above left=-1.5mm] {\footnotesize $\alpha_1$} (2_1);
\draw[->] (2_1) to[parallel segment] node[below right=-1.5mm] {\footnotesize $\beta_3$} (3_0);

\draw[<-] ($(3_0.east) + (0mm, 0.5mm)$) -- node[above=0mm] {\footnotesize $\alpha_c$} ($(3_0.east) + (5mm, 0.5mm)$);
\draw[->] ($(3_0.east) + (0mm, -0.5mm)$) -- node[below=0mm] {\footnotesize $\beta_c$} ($(3_0.east) + (5mm, -0.5mm)$);

\node [circle, below= -5.27mm of 2_1.base] (base) {};
\end{tikzpicture}}
  \quad
  \subfloat[Operations]{\label{fig:fast-ssc-tree-annotated:b}\definecolor{deepgreen}{RGB}{8, 130, 25}

\begin{tikzpicture}[baseline=(base),
        level/.style={level distance = 6mm},
        level 1/.style={sibling distance=19mm, edge from parent/.style={draw,black,line width=2pt}},
        level 2/.style={sibling distance=9mm, edge from parent/.style={draw,black,line width=1pt}},
        level 3/.style={sibling distance=4mm, edge from parent/.style={draw,black,line width=0.5pt}},
        ]

\tikzset{
frozen/.style={thick,draw=black,fill=white,minimum size=3mm,circle, inner sep=0},
fullspace/.style={thick,draw=black,fill=black,minimum size=3mm,circle, inner sep = 0},
mixed/.style={thick,draw=black,fill=gray,minimum size=3mm,circle, inner sep = 0},
rep_mixed/.style={thick,draw=black,pattern=north west lines,pattern color=deepgreen,minimum size=3mm,circle, inner sep = 0},
spc_mixed/.style={thick,draw=black,pattern=crosshatch,pattern color=orange,minimum size=3mm,circle, inner sep = 0},
repspc/.style={thick,draw=black,pattern=vertical lines,pattern color=blue,minimum size=3mm,circle, inner sep = 0}
}

\tikzset{
parallel segment/.style={
   segment distance/.store in=\segDistance,
   segment pos/.store in=\segPos,
   segment length/.store in=\segLength,
   to path={
   ($(\tikztostart)!\segPos!(\tikztotarget)!\segLength/2!(\tikztostart)!\segDistance!90:(\tikztotarget)$) -- 
   ($(\tikztostart)!\segPos!(\tikztotarget)!\segLength/2!(\tikztotarget)!\segDistance!-90:(\tikztostart)$)  \tikztonodes
   }, 
   segment pos=.5,
   segment length=4ex,
   segment distance=-1mm,
},
}

\node[mixed] (3_0){} [grow=left]
	child {node[shape=circle split, circle split part fill={white,gray},
    scale=0.625, draw=black, line width=0.275mm] (2_0){}
    child {node[fullspace,yshift=-4mm] (1_2){}
    }
	}
	child {node[spc_mixed] (2_1){}
	}
;
	
\draw[->] (3_0) to[parallel segment] node[above right=-1.5mm] (F8) {\footnotesize \scalebox{.8}{F<8>}} (2_0);
\draw[->] (2_0) to[parallel segment, segment length=2.5ex] node[above left=-1.5mm] {\footnotesize \scalebox{.8}{G\_0R<4>}} (1_2);
\draw[->] (1_2) to[parallel segment, segment length=2.5ex] node[coordinate] (I2) {} (2_0);
\node at ([xshift=4.5mm,yshift=-0.5mm]1_2) {\footnotesize \scalebox{.75}{I<2>}};

\draw[->] (2_0) to[parallel segment] node[coordinate] {} (3_0);
\node at ([xshift=-7.25mm,yshift=2.75mm]3_0) {\footnotesize \scalebox{.75}{C\_0R<4>}};

\draw[->] (3_0) to[parallel segment] node[above left=-1.5mm] {\footnotesize \scalebox{.8}{G<8>}} (2_1);
\draw[->] (2_1) to[parallel segment] node[below right=-1.5mm] {\footnotesize \scalebox{.8}{SPC<4>}} (3_0);

\draw[<-] ($(3_0.east) + (0mm, 0.5mm)$) -- node[above=0mm] {\footnotesize \scalebox{.8}{$\alpha_c$}} ($(3_0.east) + (5mm, 0.5mm)$);
\draw[->] ($(3_0.east) + (0mm, -0.5mm)$) -- node[label={[label distance=-3mm]-2mm:{\footnotesize \scalebox{.8}{C<8>}}}] {} ($(3_0.east) + (5mm, -0.5mm)$);

\end{tikzpicture}}
  \caption{Dataflow graph of a $(8,5)$ polar decoder.}
  \label{fig:fast-ssc-tree-annotated}
\end{figure}

A tool was built to generate a list of function calls corresponding to the decoder tree traversal. It was first described in \cite{Sarkis_GLOBALSIP_2014} and has been significantly improved since its initial publication notably to add support for other node types as well as to add support for GPU code generation. Listing~\ref{lst:unrolled} shows an example decoder that corresponds to the (8, 5) polar code whose dataflow graph is shown in Fig.~\ref{fig:fast-ssc-tree-annotated}. For brevity and clarity, in Fig.~\ref{fig:fast-ssc-tree-annotated:b}, I and C\_0R correspond to the Info and Combine\_0R functions, respectively.

\floatname{algorithm}{Listing}
\begin{algorithm}[t]
\caption{Unrolled (8, 5) Fast-SSC Decoder}
\label{lst:unrolled}
\begin{algorithmic}
\State F\tparam{8}($\alpha_c$, $\alpha_1$);
\State G\_0R\tparam{4}($\alpha_1$, $\alpha_2$);
\State Info\tparam{2}($\alpha_2$, $\beta_1$);
\State Combine\_0R\tparam{4}($\beta_1$, $\beta_2$);
\State G\tparam{8}($\alpha_c$, $\alpha_2$, $\beta_2$);
\State SPC\tparam{4}($\alpha_2$, $\beta_3$);
\State Combine\tparam{8}($\beta_2$, $\beta_3$, $\beta_c$);
\end{algorithmic}
\end{algorithm}

\subsubsection{Eliminating Superfluous Operations on $\beta$-Values}
Every non-leaf node in the decoder performs the combine operation \eqref{eq:combine}, rendering it the most common operation. In \eqref{eq:combine}, half the $\beta$ values are copied unchanged to $\beta_v$. One meth\-od to significantly reduce decoding latency is to eliminate those superfluous copy operations by choosing an appropriate layout for $\beta$ values in memory: Only $N$ $\beta$ values are stored in a contiguous array aligned to the SIMD vector size. When a combine operation is performed, only those values corresponding to $\beta_l$ will be updated.
Since the stage sizes are all powers of two, stages of sizes equal to or larger than the SIMD vector size will be implicitly aligned so that operations on them are vectorized.

\subsubsection{Improved Layout of the $\alpha$-memory}\label{sec:alpha_mem}
Unlike in the case of $\beta$ values, the operations producing $\alpha$ values, $f$ and $g$ operations, do not copy data unchanged. Therefore, it is important to maximize the number of vectorized operations to increase decoding speed. To this end, contiguous memory is allocated for the $\log_2 N$ stages of the decoder. The overall memory and each stage is aligned to 16 or 32-byte boundaries when SSE or AVX instructions are used, respectively. As such, it becomes possible to also vectorize stages smaller than the SIMD vector size. The memory overhead due to not tightly packing the stages of $\alpha$ memory is negligible. As an example, for an $N = \num[group-separator={,}]{32768}$ floating-point polar decoder using AVX instructions, the size of the $\alpha$ memory required by the proposed scheme is 262,208 bytes, including a 68-byte overhead.

\subsubsection{Compile-time Specialization}
Since the sizes of the constituent codes are known at compile time, they are provided as template parameters to the functions as illustrated in Listing~\ref{lst:unrolled}. Each function has two or three implementations. One is for stages smaller than the SIMD vector width where vectorization is not possible or straightforward. A second one is for stages that are equal or wider than the largest vectorization instruction set available. Finally, a third one provides SSE vectorization in an AVX or AVX2 decoder for stages that can be vectorized by the former, but are too small to be vectorized using AVX or AVX2. The last specialization was noted to improve decoding speed in spite of the switch between the two SIMD extension types.

Furthermore, since the bounds of loops are compile-time constants, the compiler is able to unroll loops where it sees fit, eliminating the remaining branches in the decoder unless they help in increasing speed by resulting in a smaller executable.

\subsubsection{Architecture-specific Optimizations}
First, the decoder was updated to take advantage of AVX2 instructions when available. These new instructions benefit the fixed-point implementation as they allow simultaneous operations on 32 8-bit integers.

Second, the implementation of some nodes were hand-optimized to better take advantage of the processor architecture. For example, the SPC node was mostly rewritten. Listing~\ref{lst:findidxmin} shows a small but critical subsection of the SPC node calculations where the index within a SIMD vector corresponding to the specified value is returned. The reduction operation required by the Repetition node has also been optimized manually.

\floatname{algorithm}{Listing}
\begin{algorithm}[t]
\caption{Finding the index of a given value in a vector}
\label{lst:findidxmin}
\small
\begin{lstlisting}[mathescape]
std::uint32_t findIdx($\alpha^*$ $x$, $\alpha$ $x_{\min}$) {
  __mm256 minVec = _mm256_broadcastb_epi8($x_{\min}$);
  __mm256 mask = _mm256_cmpeq_epi8(minVec, $x$);
  std::uint32_t mvMask = _mm256_movemask_epi8(mask);
  return __tzcnt_u32(mvMask);
}
\end{lstlisting}
\end{algorithm}

Third, for the floating-point implementation, $\beta$ was chan\-ged to be in $\{+0,-0\}$ instead of $\{+1,-1\}$. In the floating-point representation~\cite{IEEE754}, the most significant bit only carries the information about the sign. Flipping this bit effectively changes the sign of the number. By changing the mapping for $\beta$, multiplications are replaced by faster bitwise XOR operations. Similarly, for the 8-bit fixed-point implementation, $\beta$ was changed to be in $\{0, -128\}$ to reduce the complexity of the Info and G functions.

Listings~\ref{lst:gvec} and \ref{lst:gvec:int8} show the resulting G functions for both the floating-point and fixed-point implementations as examples illustrating bottom-up optimizations used in our deco\-ders.

\floatname{algorithm}{Listing}
\begin{algorithm}[t]
\caption{Vectorized floating-point G function ($g$ operation)}
\label{lst:gvec}
\small
\begin{lstlisting}[mathescape]
template<unsigned int $N_v$>
void G($\alpha^*$ $\alpha_{in}$, $\alpha^*$ $\alpha_{out}$, $\beta^*$ $\beta_{in}$) {
  for (unsigned int i = 0; i < $\nicefrac{N_v}{2}$; i += 8) {
    __m256 $\alpha_l$ = _mm256_load_ps($\alpha_{in}$ + i);
    __m256 $\alpha_r$ = _mm256_load_ps($\alpha_{in}$ + i + $\nicefrac{N_v}{2}$);
    __m256 $\beta_v$ = _mm256_load_ps($\beta_{in}$ + i);
    __m256 $\alpha_l'$ = _mm256_xor_ps($\beta_v$, $\alpha_l$);
    __m256 $\alpha_v$ = _mm256_add_ps($\alpha_r$, $\alpha_l'$);
    __mm256_store_ps($\alpha_{out}$ + i, $\alpha_v$);
  }
}
\end{lstlisting}
\end{algorithm}

\floatname{algorithm}{Listing}
\begin{algorithm}[t]
\caption{Vectorized 8-bit fixed-point G function ($g$ operation)}
\label{lst:gvec:int8}
\small
\begin{lstlisting}[mathescape]
static const __m256i ONE = _mm256_set1_epi8(1);
static const __m256i M127 = _mm256_set1_epi8(-127);

template<unsigned int $N_v$>
void G($\alpha^*$ $\alpha_{in}$, $\alpha^*$ $\alpha_{out}$, $\beta^*$ $\beta_{in}$) {
  for (unsigned int i = 0; i < $\nicefrac{N_v}{2}$; i += 32) {
    __m256i $\alpha_l$ = _mm256_load_si256($\alpha_{in}$ + i);
    __m256i $\alpha_r$ = _mm256_load_si256($\alpha_{in}$ + i + $\nicefrac{N_v}{2}$);
    __m256i $\beta_v$ = _mm256_load_si256($\beta_{in}$ + i);
    __m256i $\beta_v'$ = _mm256_or_si256($\beta_v$, ONE);
    __m256i $\alpha_l'$ = _mm256_sign_epi8($\alpha_l$, $\beta_v'$);
    __m256i $\alpha_v$ = _mm256_add_ps($\alpha_r$, $\alpha_l'$);
    __m256i $\alpha_v'$ = _mm256_max_epi8(M127, $\alpha_v$);
    __mm256_store_si256($\alpha_{out}$ + i, $\alpha_v'$);
  }
}
\end{lstlisting}
\end{algorithm}

\subsubsection{Memory Footprint}\label{sec:mem_footprint}
The memory footprint is considered an important constraint for software applications. Our proposed implementations use 2 contiguous memory blocks that correspond to the $\alpha$ and $\beta$ values, respectively. The size of the $\beta$-memory is
\begin{equation}
M_{\beta}=NW_{\beta},
\end{equation}
where $N$ is the frame length, $W_{\beta}$ is the number of bits used to store a $\beta$ value and $M_{\beta}$ is in bits.

The size of the $\alpha$-memory can be expressed as
\begin{equation}
M_{\alpha}=\left[(2N-1)+A\log_2A-\left(\sum_{i=0}^{\log_2(A)-1}2^i\right)\right]W_{\alpha},
\end{equation}
where $N$ is the frame length, $W_{\alpha}$ is the number of bits used to store an $\alpha$ value, $A$ is the number of $\alpha$ values per SIMD vector and $M_{\alpha}$ is in bits. Note that the expression of $M_{\alpha}$ contains the expression for the overhead $M_{\alpha\text{OH}}$ due to tightly packing the $\alpha$ values as described in Section~\ref{sec:alpha_mem}:
\begin{equation}
M_{\alpha\text{OH}}=\left[ A\log_2A-\left(\sum_{i=0}^{\log_2(A)-1}2^i\right)\right] W_{\alpha}.
\end{equation}

The memory footprint can thus be expressed as
\begin{equation}
\begin{split}
M_{\text{total}} &= M_{\beta}+M_{\alpha}\\
&= NW_{\beta} + \left[(2N-1)+A\log_2A-\left(\sum_{i=0}^{\log_2(A)-1}2^i\right)\right]W_{\alpha}.
\end{split}
\end{equation}
The memory footprint in kilobytes can be approximated with
\begin{equation}
M_{\text{total (kbytes)}} \approx \frac{N(W_{\beta} + 2W_{\alpha})}{8000}.
\end{equation}

\subsubsection{Implementation Comparison}
We first compare the SIMD-float results for this implemen\-ta\-tion---the unrolled decoder---with those from Section~\ref{sec:impl:ni}---the instruction-based decoder. Then we show SIMD-int8 results and compare them with that of the software decoder of Le Gal et. al \cite{LeGal_TSP_2014}. As in the previous sections, the results are for an Intel Core i7-4770S running at 3.1 GHz when Turbo is disabled and at up to 3.9 GHz otherwise. The decoders were limited to a single CPU core.

\begin{table}[t]
  \centering
  \caption{Decoding polar codes with floating-point precision using SIMD, comparing the instruction-based decoder (ID) with the unrolled decoder (UD).}
  \begin{tabular}{c c c c c}
    \toprule
    \multirow{3}{*}{\shortstack{Code\\$(N,k)$}} & \multicolumn{2}{c}{Info T/P (Mbps)} & \multicolumn{2}{c}{Latency ($\mu$s)} \\
    \cmidrule(lr){2-3} \cmidrule(lr){4-5}
                    & ID    & UD    & ID  & UD\\
    \midrule
    $(2048,1024)$   &  75.6 & 229.8 &  14 &   4\\
    $(2048,1707)$   & 173.9 & 492.2 &  10 &   3\\
    $(32768,27568)$ & 124.3 & 271.3 & 222 & 102\\
    $(32768,29492)$ & 160.1 & 315.1 & 184 &  94\\
    \bottomrule
  \end{tabular}%
  \label{tab:impl:tp:cmp23}
\end{table}

Table~\ref{tab:impl:tp:cmp23} shows the impact of the optimizations introduced in the unrolled version on the SIMD-float implementations. It resulted in the unrolled decoders being 2 to 3 times faster than the flexible, instruction-based, ones. Comparing Tables~\ref{tab:impl:tp2} and \ref{tab:impl:tp:cmp23} shows an improvement factor from 3.3 to 5.7 for the SIMD-int8 implementations. It should be noted that some of the improvements introduced in the unrolled decoders could be backported to the instruction-based decoders, and is considered for future work.

\begin{table*}[t]
  \centering
  \caption{Comparison of the proposed software decoder with that of \cite{LeGal_TSP_2014}.}
  \begin{tabular}{c c c c c c c c}
    \toprule
    \multirow{2}{*}{Decoder} & \multirow{2}{*}{Target} & \multirow{2}{*}{\shortstack{L3\\Cache}} & \multirow{2}{*}{\shortstack{$f$\\(GHz)}} & \multirow{2}{*}{\shortstack{Code\\$(N,k)$}} & \multirow{2}{*}{\shortstack{Mem. footprint\\(kbytes)}} & \multirow{2}{*}{\shortstack{Info T/P\\(Mbps)}} & \multirow{2}{*}{\shortstack{Latency\\($\mu$s)}} \\
    &&&&&&&\\
    \midrule
    \cite{LeGal_TSP_2014}*& Intel Core i7-4960HQ & 6MB & 3.6+ & $(2048,1024)$   &  144 & 1,320 & 25\\
                          &                      &     &      & $(2048,1707)$   &  144 & 2,172 & 26\\
                          &                      &     &      & $(32768,27568)$ & 2304 & 1,232 & 714\\
                          &                      &     &      & $(32768,29492)$ & 2304 & 1,557 & 605\vspace{2pt}\\
    this work\phantom{*}  & Intel Core i7-4770S  & 8MB & 3.1  & $(2048,1024)$   &    6 &   398 & 3\\
                          &                      &     &      & $(2048,1707)$   &    6 & 1,041 & 2\\
                          &                      &     &      & $(32768,27568)$ &   98 &   886 & 31\\
                          &                      &     &      & $(32768,29492)$ &   98 & 1,131 & 26\vspace{2pt}\\
    this work*            & Intel Core i7-4770S  & 8MB & 3.1+ & $(2048,1024)$   &    6 &   502 & 2\\
                          &                      &     &      & $(2048,1707)$   &    6 & 1,293 & 1\\
                          &                      &     &      & $(32768,27568)$ &   98 & 1,104 & 25\\
                          &                      &     &      & $(32768,29492)$ &   98 & 1,412 & 21\\
    \bottomrule
    &\vspace{-8pt}\\
    \multicolumn{6}{l}{*\textit{Results with Turbo enabled.}}\\
  \end{tabular}%
  \label{tab:impl:tp_vs_gal}
\end{table*}

Compared to the software polar decoders of \cite{LeGal_TSP_2014}, Table~\ref{tab:impl:tp_vs_gal} shows that our throughput is lower for short frames but can be comparable for long frames. However, latency is an order of magnitude lower for all code lengths. This is to be expected as the decoders of \cite{LeGal_TSP_2014} do inter-frame parallelism i.e. parallelize the decoding of independent frames while we parallelize the decoding of a frame. The memory footprint of our decoder is shown to be approximately 24 times lower than that of \cite{LeGal_TSP_2014}. The results in \cite{LeGal_TSP_2014} were presented with Turbo frequency boost enabled; therefore we present two sets of results for our proposed decoder: one with Turbo enabled, indicated by the asterisk (*) and the $3.1+$ GHz frequency in the table, and one with Turbo disabled. The results with Turbo disabled are more indicative of a full SDR system as all CPU cores will be fully utilized, not leaving any thermal headroom to increase the frequency. The maximum Turbo frequencies are 3.8 GHz and 3.9 GHz for the i7-4960HQ and i7-4770S CPUs, respectively.

Looking at the first two, or last two rows of Table~\ref{tab:impl:tp:cmp23}, it can be seen that for a fixed code length, the decoding latency is smaller for higher code rates. The tendency of decoding latency to decrease with increasing code rate and length was first discussed in \cite{Sarkis2013}. It was noted that higher rate codes resulted in SSC decoder trees with fewer nodes and, therefore, lower latency. Increasing the code length was observed to have a similar, but lesser, effect. However, once the code becomes sufficiently long, the limited memory bandwidth and number of processing resources form bottlenecks that negate the speed gains.

The effects of unrolling and using the Fast-SSC algorithm instead of SC are illustrated in Table~\ref{tab:impl:tp:algo-unroll}. It can be observed that unrolling the Fast-SSC decoder results in a 5 time decrease in latency. Using the Fast-SSC instead of SC decoding algorithm decreased the latency of the unrolled decoder by 3 times.

\begin{table}[t]
  \centering
  \caption{Effect of unrolling and algorithm choice on decoding speed of the (2048, 1707) code on the Intel Core i7-4770S}
  \begin{tabular}{c c c}
    \toprule
    Decoder & Info T/P (Mbps) & Latency ($\mu$s)\\
    \midrule
    ID & 210 & 8.1\\
    UD SC & 363 & 4.7\\
    UD Fast-SSC & 1041 & 1.6\\
    \bottomrule
  \end{tabular}%
  \label{tab:impl:tp:algo-unroll}
\end{table}

\section{Implementation on Embedded Processors}
\label{sec:arm}
Many of the current embedded processors used in SDR applications also offer SIMD extensions, e.g. NEON for ARM processors. All the strategies used to develop an efficient x86 implementation can be applied to the ARM architecture with changes to accommodate differences in extensions. For example, on ARM, there is no equivalent to the \texttt{movemask} SSE/AVX x86 instruction.

The equations for the memory footprint provided in Section~\ref{sec:mem_footprint} also apply to our decoder implementation for embedded processors.

\subsubsection*{Comparison with Similar Works}
Results were obtained using the ODROID-U3 board, which features a Samsung Exynos 4412 system on chip (SoC) implementing an ARM Cortex A9 clocked at 1.7 GHz. Like in the previous sections, the decoders were only allowed to use one core. Table~\ref{tab:impl:arm} shows the results for the proposed unrolled decoders and provides a comparison with \cite{LeGal_SIPS_2014}. As with their desktop CPU implementation of \cite{LeGal_TSP_2014}, inter-frame parallelism is used in the latter.

\begin{table}[t]
  \centering
  \setlength{\tabcolsep}{4pt}
  \caption{Decoding polar codes with 8-bit fixed-point numbers on an ARM Cortex A9 using NEON.}
  \begin{tabular}{c c c c c c}
    \toprule
    \multirow{3}{*}{\shortstack{Code\\$(N,k)$}} & \multirow{3}{*}{Decoder} & \multirow{3}{*}{\shortstack{Mem.\\Footprint\\\vspace{1pt}(kBytes)}} & \multicolumn{2}{c}{T/P (Mbps)} & \multirow{3}{*}{\shortstack{Latency\\($\mu$s)}}\\
    \cmidrule(lr){4-5}
                    &                                  &      & Coded & Info &       \\
    \midrule                                                  
    $(1024,512)$    & \cite{LeGal_SIPS_2014}\phantom{*}&    38&  70.5 & 35.3 &    232\\
                    & \cite{LeGal_SIPS_2014}*          &    38&  80.6 & 42.9 &    191\vspace{2pt}\\
                    & this work                        &     3& 113.1 & 56.6 &      9\vspace{4pt}\\
    $(32768,29492)$ & \cite{LeGal_SIPS_2014}\phantom{*}& 1,216&  33.1 & 29.8 & 15,844\\
                    & \cite{LeGal_SIPS_2014}*          & 1,216&  40.2 & 36.2 & 13,048\vspace{2pt}\\
                    & this work                        &    98&  90.8 & 81.7 &    361\\
    \bottomrule
    &\vspace{-8pt}\\
    \multicolumn{6}{l}{*\textit{Results linearly scaled for the clock frequency difference.}}\\
  \end{tabular}%
  \label{tab:impl:arm}
\end{table}

It can be seen that the proposed implementations provide better latency and greater throughput at native frequencies.
Since the ARM CPU in the Samsung Exynos 4412 is clocked at 1.7 GHz while that in the NVIDIA Tegra 3 used in \cite{LeGal_SIPS_2014} is clocked at 1.4 GHz, we also provide linearly scaled throughput and latency numbers for the latter work, indicated by an asterisk (*) in the table.
Compared to the scaled results of \cite{LeGal_SIPS_2014}, the proposed decoder has 1.4--2.25 times the throughput and its latency is 25--36 times lower. The memory footprint of our proposed decoder is approximately 12 times lower than that of \cite{LeGal_SIPS_2014}. Both implementations are using 8-bit fixed-point values.

\section{Implementation on Graphical Processing Units}
\label{sec:gpu}
Most recent graphical processing units (GPU) have the capability to do calculations that are not related to graphics. These GPUs are often called general purpose GPUs (GPGPU). In this section, we describe our approach to implement software polar decoders in CUDA C~\cite{nvidia2014} and present results for these decoders running on a NVIDIA Tesla K20c.

Most of the optimization strategies cited above could be applied or adapted to the GPU. However, there are noteworthy differences. Note that, when latency is mentioned below we refer to the decoding latency including the delay required to copy the data in and out of the GPU.

\subsection{Overview of the GPU Architecture and Terminology}

A NVIDIA GPU has multiple microprocessors with 32 cores each. Cores within the same microprocessor may communicate and share a local memory. However, synchronized communication between cores located in different microprocessors often has to go through the CPU and is thus costly and discouraged \cite{Feng2010}.

GPUs expose a different parallel programming model than general purpose processors. Instead of SIMD, the GPU model is single-instruction-multiple-threads (SIMT). Each core is capable of running a thread. A computational kernel performing a specific task is instantiated as a block. Each block is mapped to a microprocessor and is assigned one thread or more.

As it will be shown in Sect.~\ref{sec:gpu:blocksperkernel}, the latency induced by transferring data in and out of a GPU is high. To minimize decoding latency and maximize throughput, a combination of intra- and inter-frame parallelism is used for the GPU contrary to the CPUs where only the former was applied. We implemented a kernel that decodes a single frame. Thus, a block corresponds to a frame and attributing e.g. 10 blocks to a kernel translates into the decoding of 10 frames in parallel.

\subsection{Choosing an Appropriate Number of Threads per Block}

As stated above, a block can only be executed on one microprocessor but can be assigned many threads. However, when more than 32 threads are assigned to a block, the threads starting at 33 are queued for execution. Queued threads are executed as soon as a core is free.

\begin{figure}[t]
  \centering
  \begin{tikzpicture}

  \pgfplotsset{
    grid style = {
      dash pattern = on 0.05mm off 1mm,
      line cap = round,
      black,
      line width = 0.5pt
    },
    y axis style/.style={
      yticklabel style=#1,
      ylabel style=#1,
      y axis line style=#1,
      ytick style=#1
    }
  }

  \begin{axis}[%
    axis y line=left,%
    y axis style=red!75!black,%
    xlabel=Threads per block,%
    ylabel=Info T/P (Mbps),%
    ylabel style={yshift=-1.0em},%
    ymin=0,%
    width=\columnwidth/1.2, height=6cm, grid=major,%
    legend style={font=\footnotesize},%
    mark size=2.0pt]

    \addplot[color=red,mark=x] table[x=threads,y=tp,col sep=semicolon]
            {vecsize.1024.922_4dB_gmem.tp.txt}
    node [pos=0.38,anchor=north, text=black] {T/P}; \label{tpplot}

  \end{axis}

  \begin{axis}[%
    axis y line=right,%
    y axis style=blue!75!black,
    ylabel=decoding latency (ms),%
    ylabel style={yshift=1.0em},%
    ymin=0,%
    width=\columnwidth/1.2, height=6cm, grid=major,%
    legend style={font=\footnotesize},%
    mark size=2.0pt]

    \addplot[color=blue,mark=o] table[x=threads,y=latency,col sep=semicolon] {vecsize.1024.922_4dB_gmem.tp.txt}
    node [pos=0.24,anchor=north, text=black] {dec. latency}; \label{latencyplot}
  \end{axis}

\end{tikzpicture}
  \caption{Effect of the number of threads per block on the information throughput and decoding latency for a $(1024, 922)$ polar code where the number of blocks per kernel is 208.}
  \label{fig:gpu_threads}
\end{figure}
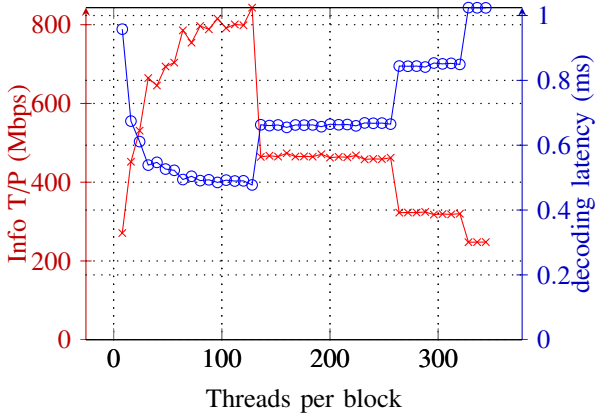

Fig.~\ref{fig:gpu_threads} shows that increasing the number of threads assigned to a block is beneficial only until a certain point is reached. For the particular case of a $(1024, 922)$ code, associating more than 128 threads to a block negatively affects performance. This is not surprising as the average node width for that code is low at 52.

\subsection{Choosing an Appropriate Number of Blocks per Kernel}
\label{sec:gpu:blocksperkernel}

Memory transfers from the host to the GPU device are of high throughput but initiating them induces a great latency. The same is also true for transfers in the other direction, from the device to the host. Thus, the number of distinct transfers have to be minimized. The easiest way to do so is to run a kernel on multiple blocks. For our application, it translates to decoding multiple frames in parallel as a kernel decodes one frame.

Yet, there s a limit to the number of resources that can be used to execute a kernel i.e. decode a frame. At some point, there will not be enough computing resources to do the work in one pass and many passes will be required. The NVIDIA Tesla K20c card features the Kepler GK110 GPU that has 13 microprocessors with 32 cores and 16 load and store units each~\cite{nvidiaGK110}. In total, 416 arithmetic or logic operations and 208 load or store operations can occur simultaneously.

Yet, there is a limit to the number of resources that can be used to execute a kernel i.e. decode a frame. At some point, there will not be enough computing resources to do the work in one pass and many passes will be required. The NVIDIA Tesla K20c card features the Kepler GK110 GPU that has 13 microprocessors with 32 cores and 16 load and store units each~\cite{nvidiaGK110}. In total, 416 arithmetic or logic operations and 208 load or store operations can occur simultaneously.

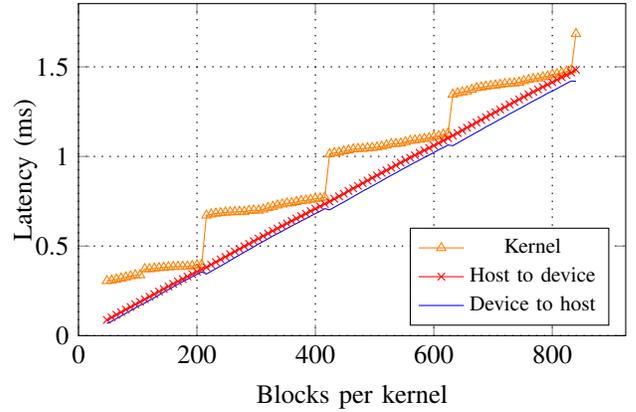
\begin{figure}[t]
  \centering
  \begin{tikzpicture}

  \pgfplotsset{
    grid style = {
      dash pattern = on 0.05mm off 1mm,
      line cap = round,
      black,
      line width = 0.5pt
    }
  }

  \begin{axis}[%
    xlabel=Blocks per kernel,%
    ylabel=Latency (ms),%
    ylabel style={yshift=-1.5em},%
    ymin=0,xmin=0,%
    width=\columnwidth, height=6cm, grid=major,%
    legend style={font=\footnotesize},%
    legend pos={south east},%
    mark size=2.0pt]

    \addplot[color=orange,mark=triangle] table[x=blocks,y expr=\thisrow{dec}*1000,col sep=semicolon]
    {mem_tx.2048.1707_3dB_gmem_t128.txt};
    \addlegendentry{Kernel}

    \addplot[color=red,mark=x] table[x=blocks,y expr=\thisrow{htod}*1000,col sep=semicolon] 
    {mem_tx.2048.1707_3dB_gmem_t128.txt};
    \addlegendentry{Host to device}

    \addplot[color=blue] table[x=blocks,y expr=\thisrow{dtoh}*1000,col sep=semicolon]
    {mem_tx.2048.1707_3dB_gmem_t128.txt};
    \addlegendentry{Device to host}

  \end{axis}

\end{tikzpicture}
  \vspace{-0.25cm}
  \caption{Effect of the number of blocks per kernel on the data transfer and kernel execution latencies for a $(2048,1707)$ polar code where the number of threads per block is 128.}
  \label{fig:gpu_mem_tx}
\end{figure}

Fig.~\ref{fig:gpu_mem_tx} shows the latency to execute a kernel, to transfer memory from the host to the GPU and vice versa for a given number of blocks per kernel. The number of threads assigned per block is fixed to 128 and the decoder is built for a $(2048, 1707)$ polar code. It can be seen that the latency of memory transfers grows linearly with the number of blocks per kernel. The kernel latency however has local minimums at multiples of 208. We conclude that the minimal decoding latency, the sum of all three latencies illustrated in Fig.~\ref{fig:gpu_mem_tx}, is bounded by the number of load and store units.

\subsection{On the Constituent Codes Implemented}

Not all the constituent codes supported by the general purpose processors are beneficial to a GPU implementation. In a SIMT model, reduction operations are costly. Moreover, if a conditional execution leads to unbalanced threads, performance suffers. Consequently, all nodes based on the single-parity-check (SPC) codes, that features both characteristics, are not used in the GPU implementation.

Experiments have shown that implementing the SPC no\-de results in a throughput reduction by a factor of 2 or more.

\subsection{Shared Memory and Memory Coalescing}

Each microprocessor contains shared memory that can be used by all threads in the same block. The NVIDIA Tesla K20c has 48 kB of shared memory per block. Individual reads and writes to the shared memory are much faster than accessing the global memory. Thus, intuitively, when conducting the calculations within a kernel, it seems preferable to use the shared memory as much as possible in place of the global memory.

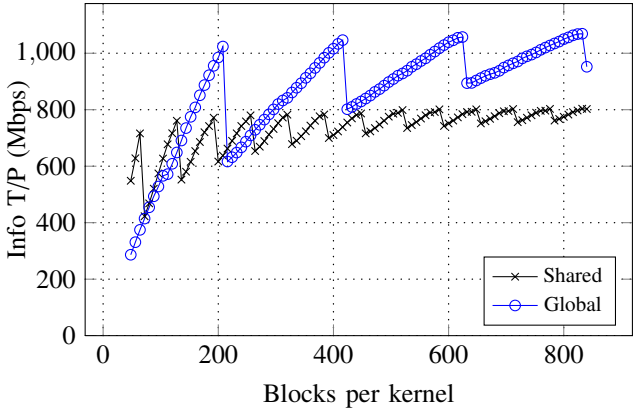
\begin{figure}[t]
  \centering
  \begin{tikzpicture}

  \pgfplotsset{
    grid style = {
      dash pattern = on 0.05mm off 1mm,
      line cap = round,
      black,
      line width = 0.5pt
    }
  }

  \begin{axis}[%
    xlabel=Blocks per kernel,%
    ylabel=Info T/P (Mbps),%
    ylabel style={yshift=-1.0em},%
    ymin=0,%
    ytick={0,200,...,1000},%
    width=\columnwidth, height=6cm, grid=major,%
    legend style={font=\footnotesize},%
    legend pos={south east},%
    mark size=2.0pt]

    \addplot[color=black,mark=x] table[x=blocks,y=tp,col sep=semicolon] 
{block.1024.922_4dB_m2_ssc+rep+p01_t128.tp.txt};
    \addlegendentry{Shared}

    \addplot[color=blue,mark=o] table[x=blocks,y=tp,col sep=semicolon] {mem_tx.1024.922_4dB_gmem_t128.txt};
    \addlegendentry{Global}

  \end{axis}
\end{tikzpicture}
  \vspace{-0.25cm}
  \caption{Information throughput comparison for a $(1024,922)$ polar code where intermediate results are stored in shared or global memory. The number of threads per block is 128.}
  \label{fig:gpu_mem_type}
\end{figure}

However, as shown by Fig.~\ref{fig:gpu_mem_type}, it is not always the case. When the number of blocks per kernel is small, using the shared memory provides a significant speedup. In fact, with 64 blocks per kernel, using shared memory results in a decoder that has more than twice the throughput compared to a kernel that only uses the global memory. Past a certain value of blocks per kernel though, solely using the global memory is clearly advantageous for our application.

These results suggest that the GPU is able to efficiently schedule memory transfers when the number of blocks per kernel is sufficiently high.

\subsection{Asynchronous Memory Transfers and Multiple Streams}

Transferring memory from the host to the device and vice versa induces a latency that can be equal to the execution of a kernel. Fortunately, that latency can be first reduced by allocating pinned or page-locked host memory. As page-locked memory can be mapped into the address space of the device, the need for a staging memory is eliminated \cite{nvidia2014}.

More significantly, NVIDIA GPUs with compute capability of 2.0 or above are able to transfer memory in and out of the device asynchronously. By creating three streams---sequences of operations that get executed in issue-order on the GPU---memory transfers and execution of the kernel can be overlapped, effectively multiplying throughput by a factor of 3.

This also increases the memory footprint by a factor of three. On the GPU, the memory footprint is
\begin{equation}
M_{\text{total (kbytes)}} = \frac{N(W_{\beta} + W_{\alpha})BS}{8000},
\end{equation}
where $B$ is the number of blocks per kernel---i.e. the number of frames being decoded simultaneously---, $S$ is the number of streams, and where $W_{\beta}$ and $W_{\alpha}$ are the number of bits required to store a $\beta$ and an $\alpha$ value, respectively. For best performance, as detailed in the next section, both $\beta$ and $\alpha$ values are represented with floating-point values and thus $W_{\beta}=W_{\alpha}=32$.

\subsection{On the Use of Fixed-Point Numbers on a GPU}

It is tempting to move calculations to 8-bit fixed-point numbers in order to speedup performance, just like we did with the other processors. However, GPUs are not optimized for calculations with integers. Current GPUs only support 32-bit integers. Even so, the maximum number of operations per clock cycle per multiprocessor as documented by NVIDIA \cite{nvidia2014} clearly shows that integers are third class citizens behind single- and double-precision floating-point numbers. As an example, Table 2 of \cite{nvidia2014} shows that GPUs with compute capability 3.5---like the Tesla K20c---can execute twice as many double-precision floating-point multiplications in a given time than it can with 32-bit integers. The same GPU can carry on 6 times more floating-point precision multiplications than its 32-bit integer counterpart.

\subsection{Results}

Table~\ref{tab:impl:gputp} shows the estimated information throughput and measured latency obtained by decoding various polar codes on a GPU. The throughput is estimated by assuming that the total memory transfer latencies are twice the latency of the decoding. This has been verified to be a reasonable assumption, using NVIDIA's profiler tool, when the number of blocks maximizes throughput.

\begin{table}[t]
  \centering
  \caption{Decoding polar codes on an NVIDIA Tesla K20c.}
  \begin{tabular}{c c c c}
    \toprule
    \multirow{2}{*}{\shortstack{Code\\$(N,k)$}} & \multirow{2}{*}{\shortstack{Nbr of\\Blocks}} & \multirow{2}{*}{\shortstack{Info T/P\\(Mbps)}} & \multirow{2}{*}{\shortstack{Latency\\(ms)}}\\
    &\\
    \midrule
    $(1024,922)$    & 208 & 1,022 & 0.6\\
                    & 416 & 1,046 & 1.1\\
                    & 624 & 1,060 & 1.6\\
                    & 832 & 1,070 & 2.2\vspace{2pt}\\
    $(2048,1707)$   & 208 &   915 & 1.1\\
                    & 416 &   936 & 2.2\\
                    & 624 &   953 & 3.3\\
                    & 832 &   964 & 4.5\vspace{2pt}\\
    $(4096,3686)$   & 208 &   959 & 2.6\\
                    & 416 & 1,002 & 4.9\\
                    & 624 & 1,026 & 6.9\\
                    & 832 & 1,043 & 9.4\\
    \bottomrule
  \end{tabular}%
  \label{tab:impl:gputp}
\end{table}

\begin{table*}[t]
  \centering
  \caption{Comparison of the power consumption and energy per information bit for the $(2048, 1707)$ polar code.}
  \begin{tabular}{c c c c c c c}
    \toprule
    \multirow{2}{*}{Decoder} & \multirow{2}{*}{Target} & \multirow{2}{*}{\shortstack{Mem. Footprint\\(kbytes)}} & \multirow{2}{*}{\shortstack{Info. T/P\\(Gbps)}} & \multirow{2}{*}{\shortstack{Latency\\($\mu$s)}} & \multirow{2}{*}{\shortstack{Power\\(W)}} & \multirow{2}{*}{\shortstack{Energy\\(nJ/info. bit)}} \\
    &\\
    \midrule
    \cite{LeGal_TSP_2014} & Intel Core i7-4960HQ*&  144 & 2.2 &   26 &  13 &   6\vspace{2pt}\\
    this work             & Intel Core i7-4770S\phantom{*} &6& 1.0 &    2 &   3 &   3\\ 
                          & Intel Core i7-4770S* &    6 & 1.3 &    1 &   5 &   4\\ 
                          & ARM Cortex A9        &    6 & 0.1 &   14 & 0.8 & 7\\ 
                          & NVIDIA Tesla K20c    &\phantom{$^{\dagger}$}3,408$^{\dagger}$ & 0.9 & 1100 & 108 & 118\\
    \bottomrule
    &\vspace{-8pt}\\
    \multicolumn{6}{l}{*\textit{Results with Turbo enabled.}}\\
    \multicolumn{6}{l}{$^{\dagger}$\textit{Amount required per stream. Three streams are required to sustain this throughput.}}\\
  \end{tabular}%
  \label{tab:impl:power_vs_gal}
\end{table*}

Performing linear regression on the results of Table~\ref{tab:impl:gputp} indicates that the latency scales linearly with the number of blocks, leading to standard error values of 0.04, 0.04 and 0.14 for the $(1024, 922)$, $(2048, 1707)$ and $(4096, 3686)$ polar codes, respectively. In our decoder, a block corresponds to the decoding a single frame. The frames are independent of each other, and so are blocks. Thus, our decoder scales well with the number of available cores.

Furthermore, looking at Table~\ref{tab:impl:gputp} it can be seen that the information throughput is in the vicinity of a gigabit per second. Experiments have shown that the execution of two kernels can slightly overlap, making our throughput results of Table~\ref{tab:impl:gputp} worst-case estimations. For example, while the information throughput to decode 832 frames of a $(4096,3686)$ polar code is estimated at 1,043 Mbps in Table~\ref{tab:impl:gputp}, the measured average value in NVIDIA's profiler tool was 1,228~Mbps, a 18\% improvement over the estimated throughput.

Our experiments have also shown that our decoders are bound by the data transfer speed that this test system is capable of. The PCIe 2.0 standard \cite{PCIe2.0} specifies a peak data throughput of 64~Gbps when 16 lanes are used and once 8b10b encoding is accounted for. Decoding 832 frames of a polar code of length $N=4096$ requires the transfer of 3,407,872 LLRs expressed as 32-bit floating-point numbers for a total of approximately 109 Mbits. Without doing any computation on the GPU, our benchmarks measured an average PCIe throughput of 45~Gbps to transfer blocks of data of that size from the host to the device and back. Running multiple streams and performing calculations on the GPU caused the PCIe throughput to drop to 40~Gbps. This corresponds to 1.25 Gbps when 32-bit floats are used to represent LLR inputs and estimated-bit outputs of the decoder. In light of these results, we conjecture that the coded throughput will remain approximately the same for any polar code as the PCIe link is saturated and data transfer is the bottleneck.

\section{Energy Consumption Comparison}
\label{sec:energy}
In this section the energy consumption is compared for all three processor types: the desktop processor, the embedded processor and the GPU. Unfortunately the Samsung Exynos 4412 SoC does not feature sensors allowing for power usage measurements of the ARM processor cores.
The energy consumption of the ARM processor was estimated from board-level measurements. An Agilent E3631A DC power supply was used to provide the 5V input to the ODROID-U3 board and the current as reported by the power supply was used to calculated the power usage when the processor was idle and under load.

On recent Intel processors, power usage can be calculated by accessing the Running Average Power Limit (RAPL) counters. The LIKWID tool suite \cite{Treibig2010} is used to measure the power usage of the processor. Numbers are for the whole processor including the DRAM package. Recent NVIDIA GPUs also feature on-chip sensors enabling power usage measurement. Steady state values are read in real-time using the NVIDIA Management Libray (NVML) \cite{nvml2014}.

Table~\ref{tab:impl:power_vs_gal} compares the energy per information bit required to decode the $(2048, 1707)$ polar code. The SIMD-int8 implementation of our unrolled decoder is compared with that of the implementation in \cite{LeGal_TSP_2014}. The former uses an Intel Core i7-4770S clocked at 3.1 GHz. The latter uses an Intel Core i7-4960HQ clocked at 3.6 GHz with Turbo enabled. The results for the ARM Cortex A9 embedded processor and NVIDIA Tesla K20c GPU are also included for comparison. Note that the GPU represents LLRs with floating-point numbers.

The energy per information bit is calculated with
\[
\text{E }(J/\text{info. bit}) = \frac{\text{P }(W)}{\text{info. T/P }(bits/s)}\,.
\]

It can be seen that the proposed decoder is slightly more energy efficient on a desktop processor compared to that of \cite{LeGal_TSP_2014}. For that polar code, the latter offers twice the throughput but at the cost of a latency that is at least 13 times greater. However, the latter is twice as fast for that polar code. Decoding on the embedded processor offers very similar energy efficiency compared to the Intel processor although the data throughput is an order of magnitude slower. However, decoding on a GPU is significantly less energy efficient than any of the decoders running on a desktop processor.

The power consumption on the embedded platform was measured to be fairly stable with only a 0.1~W difference between the decoding of polar codes of lengths 1024 or 32,768.

\section{Further Discussion}
\subsection{On the relevance of the instruction-based decoders}
Some applications require excellent error-correction performance that necessitates the use of polar codes much longer than $N = \num[group-separator={,}]{32768}$. For example, Quantum Key Distribution benefits from frames of $2^{21}$ to $2^{24}$ bits \cite{Jouguet2014}. At such lengths, current compilers fail to compile an unrolled decoder. However, the instruction-based decoders are very suitable and are capable of throughput greater than 100 Mbps with a code of length 1 million.

\subsection{On the relevance of software decoders in comparison to hardware decoders}
The software decoders we have presented are good for systems that require moderate throughput without incurring the cost of dedicated hardware solutions. For example, in a soft\-ware-defined radio communication chain based on USRP radios and the GNU Radio software framework, a forward error-correction (FEC) solution using our proposed decoders only consumes 5\% of the total execution time on the receiver. Thus, freeing FPGA resources to implement functions other than FEC, e.g. synchronization and demodulation.

\subsection{Comparison with LDPC codes}
LDPC codes are in widespread use in wireless communication systems. In this section, the error-correction performance of moderate-length polar codes is compared against that of standard LDPC codes \cite{802.11n}. Similarly, the performance of the state-of-the-art software LDPC decoders is compared against that of our proposed unrolled decoders for polar codes.

The fastest software LDPC decoders in literature are tho\-se
of \cite{Han2013}, which implements decoders for the 802.11n standard and present results for the Intel Core i7-2600 x86 processor. That wireless communication standard defines three code lengths: 1944, 1296, 648; and four code rates: 1/2, 2/3, 3/4, 5/6. In \cite{Han2013}, LDPC decoders are implemented for all four codes rates with a code length of 1944. A layered offset-min-sum decoding algorithm with five iterations is used and early-termination is not supported.

\begin{figure}[t]
  \centering
  \definecolor{darkgreen}{RGB}{0,128,0}

\begin{tikzpicture}
  \pgfplotsset{
    grid style = {
      dash pattern = on 0.05mm off 1mm,
      line cap = round,
      black,
      line width = 0.5pt
    },
    label style = {font=\fontsize{9pt}{7.2}\selectfont},
    tick label style = {font=\fontsize{8pt}{7.2}\selectfont}
  }

  \begin{semilogyaxis}[%
    xlabel=$E_b/N_0$ (dB),xtick={1.5,2,2.5,3,3.5,4,4.5,5},%
    xlabel style={yshift=0.6em},%
    ylabel=Frame-error rate, ylabel style={yshift=-1.0em},%
    ymin=5e-5,%
    width=0.85\columnwidth, height=6cm, grid=major,%
    legend style={
      anchor={center},
      cells={anchor=west},
      column sep=0.75mm,
      font={\fontsize{8pt}{0.2}\selectfont}
    },
    legend columns=5,
    legend to name=vs-wifi-legend,
    mark size=3.0pt]

    \addlegendimage{empty legend}
    \addlegendentry{\textbf{Polar:}}

    \addplot[color=black,densely dashed] table[x=snr_db,y=FER]{2k5.s0.708.int8.txt};
    \addlegendentry{R=1/2}

    \addplot[color=blue,mark=triangle,densely dashed] table[x=ebn0_db,y=FER]{2k_r23_s0.596.txt};
    \addlegendentry{R=2/3}

    \addplot[color=darkgreen,mark=o,densely dashed] table[x=ebn0_db,y=FER]{2k_r34_s0.530.txt};
    \addlegendentry{R=3/4}

    \addplot[color=red,densely dashed,mark=star] table[x=ebn0_db,y=FER]{2k_r56_s0.461.txt};
    \addlegendentry{R=5/6}

    \addlegendimage{empty legend}
    \addlegendentry{\textbf{LDPC:}}

    \addplot[color=black] table[x=ebn0_db,y=FER]{wifi_1944_r12.txt};
    \addlegendentry{R=1/2}

    \addplot[color=blue,mark=triangle] table[x=ebn0_db,y=FER]{wifi_1944_r23.txt};
    \addlegendentry{R=2/3}

    \addplot[color=darkgreen,mark=o] table[x=ebn0_db,y=FER]{wifi_1944_r34.txt};
    \addlegendentry{R=3/4}

    \addplot[color=red,mark=star] table[x=ebn0_db,y=FER]{wifi_1944_r56.txt};
    \addlegendentry{R=5/6}

 \end{semilogyaxis}
\end{tikzpicture}
\\
\ref{vs-wifi-legend}
  \caption{Error-correction performance of the polar codes of length 2048 compared with the LDPC codes of length 1944 from the 802.11n standard. }
  \label{fig:vs_wifi}
\end{figure}
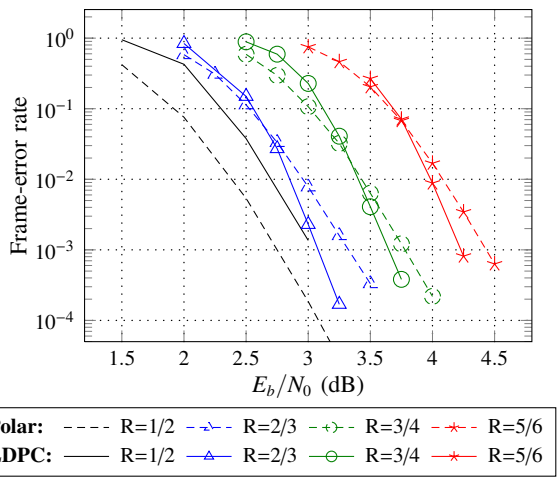

Fig.~\ref{fig:vs_wifi} shows the frame-error rate (FER) of these codes using 10 iterations of a flooding-schedule offset min-sum floating-point decoding algorithm which yields slightly better results than the five iteration layered algorithm used in \cite{Han2013}. The FER of polar codes with a slightly longer length of 2048 and matching code rates are also shown in Fig.~\ref{fig:vs_wifi}.

Table~\ref{tab:wifi-speed} that provides the latency and information throughput for decoding 524,280 information bits using the state-of-the-art software LDPC decoders of \cite{Han2013} compared to our proposed polar decoders. To remain consistent with the result presented in \cite{Han2013}, which used the Intel Core i7-2600 processor, the results in Table~\ref{tab:wifi-speed} use that processor as well.

While the polar code with rate $\nicefrac{1}{2}$ offers a better coding gain than its LDPC counterpart, all other polar codes in Fig.~\ref{fig:vs_wifi} are shown to suffer a coding loss close to 0.25 dB at a FER of $10^{-3}$. However, as Table~\ref{tab:wifi-speed} shows, there is approximately an order of magnitude advantage for the proposed unrolled polar decoders in terms of both latency and throughput compared to the LDPC decoders of \cite{Han2013}.

\begin{table}[t]
  \caption{Information throughput and latency of the polar decoders compared with the LDPC decoders of \cite{Han2013} when estimating 524,280 information bits on a Intel Core i7-2600.}
  \setlength{\tabcolsep}{0.175cm}
  \begin{tabular}{c c c c c c}
    \toprule
    \multirow{3}{*}{Decoder} & \multirow{3}{*}{$N$} & \multirow{3}{*}{Rate} & \multicolumn{2}{c}{Latency} & \multirow{3}{*}{\shortstack{Info. T/P\\(Mbps)}}\\
    \cmidrule{4-5}
    & & & total (ms) & per frame ($\mu$s) &\\
    \midrule
    \cite{Han2013} & 1944 & 1/2 & 17.4 & N/A & 30.1\\
                   &      & 2/3 & 12.7 & N/A & 41.0\\
                   &      & 3/4 & 11.2 & N/A & 46.6\\
                   &      & 5/6 & 9.3  & N/A & 56.4\vspace{1.2pt}\\
    this work      & 2048 & 1/2 & 2.0 & 3.83 & 267.4 \\
                   &      & 2/3 & 1.0 & 2.69 & 507.4 \\
                   &      & 3/4 & 0.8 & 2.48 & 619.4 \\
                   &      & 5/6 & 0.6 & 2.03 & 840.9 \\
    \bottomrule
  \end{tabular}
  \label{tab:wifi-speed}
\end{table}

\section{Conclusion}
\label{sec:conclusion}
In this work, we presented low-latency software polar decoders adapted to different processor architectures. The decoding algorithm is adapted to exploit different SIMD instruction sets for the desktop and embedded processors (SSE, AVX and NEON) or to the SIMT model inherent to GPU processors. The optimization strategies go beyond parallelisation with SIMD or SIMT. Most notably, we proposed to generate a branchless fully unrolled decoder, to use compile-time specialization, and adopt a bottom-up approach by adapting the decoding algorithm and data representation to features offered by processor architectures. For desktop processors, we have shown that intra-frame parallelism can be exploited to get a very low-latency while achieving information throughputs greater than 1 Gbps using a single core. For embedded processors, the principle remains but the achievable information throughputs are more modest at 80 Mbps. On the GPU we showed that inter-frame parallelism could be successfully used in addition to intra-frame parallelism to reach better speed, and the impact of two critical parameters on the performance of the decoders was explored. We showed that given the right set of parameters, GPU decoders are able to sustain an information throughput around 1 Gbps while simultaneously decoding hundreds of frames. Finally, we showed that the memory footprint of our proposed decoder is at least an order of magnitude lower than that our the state-of-the-art polar decoder while being slightly more energy efficient. These results indicate that the proposed software decoders make polar codes interesting candidates for software-defined radio applications.

\section*{ACKNOWLEDGEMENT}
The authors wish to thank Samuel Gagn\'{e} of \'{E}cole de technologie sup\'{e}rieure and CMC Microsystems for providing access to the Intel Core i7-4770S processor and NVIDIA Tesla K20c graphical processing unit, respectively. Claude Thibeault is a member of ReSMiQ. Warren J. Gross is a member of ReSMiQ and SYTACom.

\bibliographystyle{IEEEtran}
\bibliography{IEEEabrv,refs}

\begin{IEEEbiography}[{\includegraphics[width=1in,height=1.25in,clip,keepaspectratio]{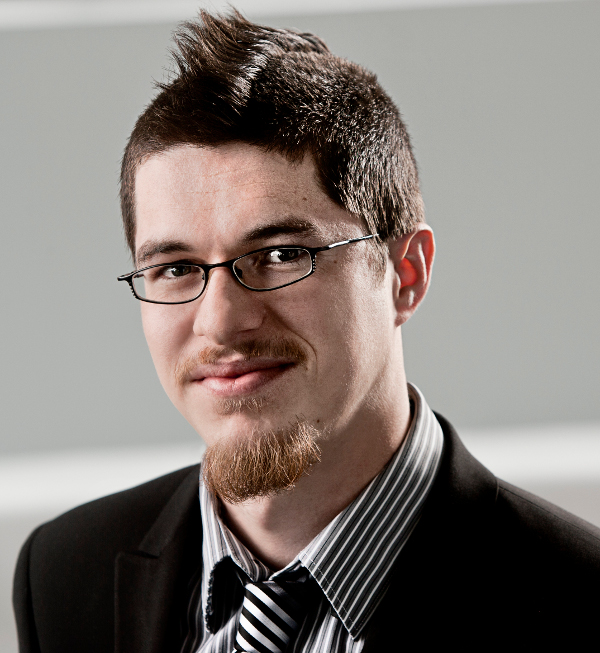}}]{Pascal Giard} received the B.Eng. and M.Eng. degree in electrical engineering from \'{E}cole de technologie sup\'erieure (\'{E}TS), Montreal, QC, Canada, in 2006 and 2009.
From 2009 to 2010, he worked as a research professional in the NSERC-Ultra Electronics Chair on 'Wireless Emergency and Tactical
Communication' at \'{E}TS.
He is currently working toward the Ph.D. degree at McGill University.
His research interests are in the design and implementation of signal processing systems with a focus on modern error-correcting codes.
\end{IEEEbiography}

\begin{IEEEbiography}[{\includegraphics[width=1in,height=1.25in,clip,keepaspectratio]{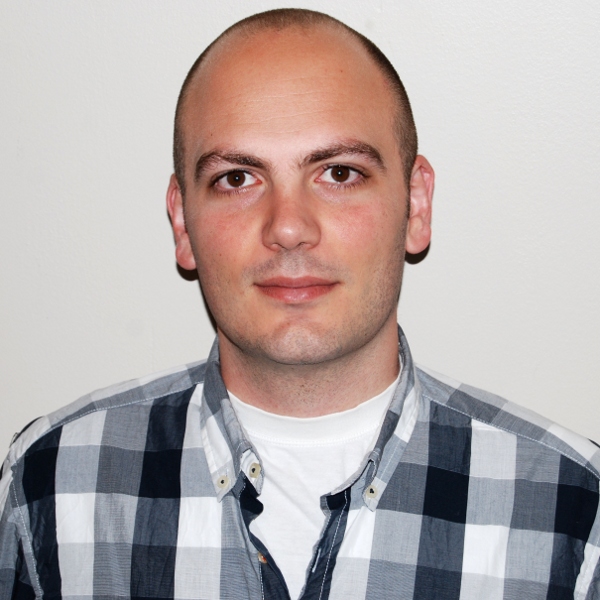}}]{Gabi Sarkis} received the B.Sc. degree in electrical engineering from Purdue University, West Lafayette, Indiana, United States, in 2006 and the M.Eng. and Ph.D. degrees from McGill University, Montreal, Quebec, Canada, in 2009 and 2016, respectively. His research interests are in the design of efficient algorithms and implementations for decoding error-correcting codes, in particular non-binary LDPC and polar codes.
\end{IEEEbiography}

\begin{IEEEbiography}[{\includegraphics[width=1in,height=1.25in,clip,keepaspectratio]{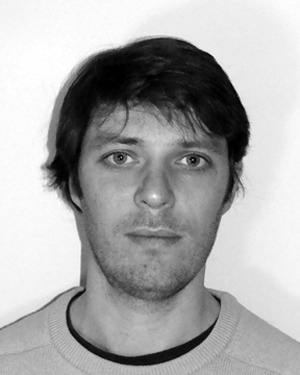}}]{Camille Leroux} received his M.Sc. degree in Electronics Engineering from the University of South Brittany, Lorient, France, in 2005. He received his Ph.D. degree in Electronics Engineering from TELECOM Bretagne, Brest, France, in 2008. From 2008 to 2011 he worked as a Post Doctoral Research Associate in the Electrical and Computer Engineering Department at McGill University, Montreal, QC, Canada. He is an Associate Professor at Bordeaux INP since 2011. His research interests are in the design and hardware implementation of telecommunication systems and computer architecture.
\end{IEEEbiography}

\begin{IEEEbiography}[{\includegraphics[width=1in,height=1.25in,clip,keepaspectratio]{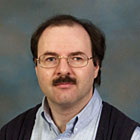}}]{Claude Thibeault}
received his Ph.D. from Ecole Polytechnique de Montreal, Canada. He is now with the Electrical Engineering department of Ecole de technologie superieure, where he serves as full professor.
His research interests include design and verification methodologies targeting ASICs and FPGAs, defect and fault tolerance, radiation effects, as well as IC and PCB test and diagnosis. He holds 13 US patents and has published more than 140 journal and conference papers, which were cited more than 850 times. He co-authored the best paper award at DVCON'05, verification category. He has been a member of different conference program committees, including the VLSI Test Symposium, for which he was program chair in 2010--2012, and general chair in 2014 and 2015.
\end{IEEEbiography}

\begin{IEEEbiography}[{\includegraphics[width=1in,height=1.25in,clip,keepaspectratio]{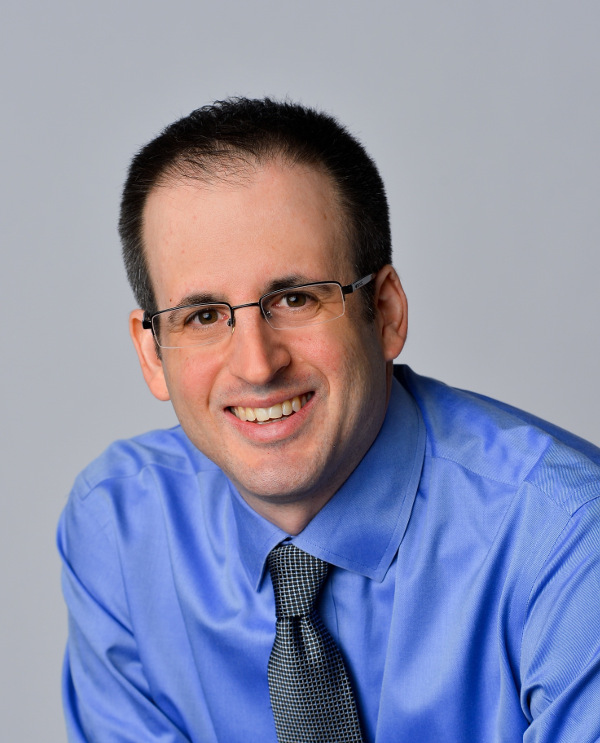}}]{Warren J. Gross}
received the B.A.Sc. degree in electrical engineering from the University of Waterloo, Waterloo, Ontario, Canada, in 1996, and the M.A.Sc. and Ph.D. degrees from the University of Toronto, Toronto, Ontario, Canada, in 1999 and 2003, respectively. Currently, he is an Associate Professor with the Department of Electrical and Computer Engineering, McGill University, Montr\'eal, Qu\'ebec, Canada. His research interests are in the design and implementation of signal processing systems and custom computer architectures.
Dr. Gross is currently Chair of the IEEE Signal Processing Society Technical Committee on Design and Implementation of Signal Processing Systems. He has served as Technical Program Co-Chair of the IEEE Workshop on Signal Processing Systems (SiPS 2012) and as Chair of the IEEE ICC 2012 Workshop on Emerging Data Storage Technologies. Dr. Gross served as Associate Editor for the IEEE Transactions on Signal Processing. He has served on the Program Committees of the IEEE Workshop on Signal Processing Systems, the IEEE Symposium on Field-Programmable Custom Computing Machines, the International Conference on Field-Programmable Logic and Applications and as the General Chair of the 6th Annual Analog Decoding Workshop. Dr. Gross is a Senior Member of the IEEE and a licensed Professional Engineer in the Province of Ontario.
\end{IEEEbiography}

\vfill 

\end{document}